\documentclass[twocolumn,showpacs,showkeys,preprintnumbers,amsmath,amssymb]{revtex4}

\usepackage{graphicx}

\usepackage{dcolumn}

\usepackage{bm}

\begin{document}



\title{ Microscopic dynamical description of proton-induced fission \\
        with the  Constrained Molecular Dynamics (CoMD) Model  }


\author{N. Vonta$^{1}$}

\author{G.A. Souliotis$^{1}$}

\email[Corresponding author. Email: ]{soulioti@chem.uoa.gr}

\author{M. Veselsky$^{2}$}


\author{A. Bonasera$^{3,4}$}


\affiliation{ $^{1}$ Laboratory of Physical Chemistry, Department of Chemistry, National and Kapodistrian \\
                     University of Athens, Athens 15771, Greece }

\affiliation{ $^{2}$Institute of Physics, Slovak Academy of
                    Sciences, Bratislava 84511, Slovakia}

\affiliation{ $^{3}$Cyclotron Institute, Texas A\&M University,
                    College Station, Texas 77843, USA}

\affiliation{ $^{4}$ Laboratori Nazionali del Sud, INFN, via Santa Sofia 62, I-95123 Catania, Italy }


\date{\today}


\begin{abstract}

The microscopic description of nuclear fission still remains a topic of intense 
basic research.  Understanding nuclear fission, apart from a theoretical point of view, 
is of practical importance for energy production and the transmutation of nuclear waste.
In nuclear astrophysics, fission sets the upper limit to the nucleosynthesis of heavy elements
via the r-process.
In this work we initiated a systematic study of intermediate energy proton-induced fission
using the Constrained Molecular Dynamics (CoMD) code. 
The CoMD code implements an effective interaction with a nuclear matter compressibility of K=200 (soft EOS) 
with several forms of the density dependence of the nucleon-nucleon symmetry potential. 
Moreover, a constraint is imposed in the phase-space occupation for each nucleon 
restoring the Pauli principle at each time step of the collision. 
A proper choice of the surface parameter of the effective interaction has been made 
to describe fission. 
In this work, we present results of fission calculations for proton-induced reactions on : 
a) $^{232}$Th at 27 and 63 MeV, 
b) $^{235}$U at 10, 30, 60 and 100 MeV, and 
c) $^{238}$U at 100 and 660 MeV.  
The calculated observables include fission-fragment mass distributions, total fission energies, 
neutron multiplicities and fission times. These observables are compared to available experimental
data.
We show that the microscopic CoMD code is able to describe the complicated many-body dynamics of the
fission process at intermediate and high energy and give a reasonable estimate of the fission time scale.
Sensitivity of the results to the density dependence of the nucleon symmetry potential 
(and, thus, the nuclear symmetry energy) is found.
Further improvements of the code are  necessary to achieve a satisfactory description of low energy fission
in which shell effects play a dominant role. 


\end{abstract}




\maketitle

\section{Introduction}

Nuclear fission, since its discovery more than 75 years ago, has played a prominent role in
applications as well as basic nuclear research.
Among the wide range of modern applications and given the increasing energy demand worldwide
\cite{oiltip}, nuclear energy production in Generation-IV reactors \cite{Bouchard-2007,NEA-2008} 
and the incineration of nuclear waste in Accelerator-Driven Systems (ADS) \cite{Aliberti-2004,Koning-2005}
are areas of current intense efforts. In parallel, fission offers an important mechanism to produce 
a variety of isotopes for medical and industrial use \cite{medical-2014}. It is also one of the 
main approaches to produce exotic neutron-rich nuclei in Rare Isotope Beam (RIB) facilities
\cite{RIBreview-2013,ATLAS,EURISOL,EURISOL-2007,RISP,RISP-2013}.

From an astrophysical point of view, fission is a key reaction of the rapid neutron capture process 
(r-process) and essentially sets the upper boundary on the synthesis of heavy elements
\cite{Goriely-2015,Goriely-2013,Panov-2013,Erler-2012,Thoenn-2011}. In the same vein, fission largely determines the  
stability and the properties of superheavy elements \cite{Oganessian-2015,SHE-2014,Thoenn-2013,Loveland-1990}.


Understanding the mechanism of nuclear fission, that is the transformation of a single heavy nucleus
into two receeding fragments, has been a long journey of fruitful research and debate and, still today, 
is far from being complete.
Upon its discovery, fission was interpreted by Meitner and Frisch \cite{Meitner-1939} as the division 
of a charged liquid drop due to the interplay of the repulsive Coulomb force between the protons and
the surface tension due to the attractive nucleon-nucleon interaction. 
In the seminal paper of Bohr and Wheeler \cite{Bohr-1939}, fission was described with a liquid-drop model
and the first estimates of fission probalilities were obtained based on statistical arguments.
The first detailed calculations of potential energies of deformed nuclear drops were performed by 
Frankel and Metropolis in 1947 \cite{Frankel-1947} employing the ENIAC, one of the first digital
computers.
Despite the success in the interpretation of fission based on the liquid-drop model, 
the prevailing asymmetry in the mass distribution of the minor actinides could not be deciphered
until shell corrections to the macroscopic liquid drop were taken into account (see below).

A detailed statistical model that could describe asymmetric fission  was developed 
by Fong \cite{Fong-1956}.     
Further advancements lead to the scission model of Wilkins \cite{Wilkins-1976} and the random 
neck-rupture model of Brosa \cite{Brosa-1990}. 
A current version of the latter is the temperature-dependent Brosa model  developed 
in \cite{Dui-2001}.  These models offer main ingredients in current statistical models
of fission (e.g. \cite{Demetriou-2010,Ryzhov-2011,Rubchenya-2012,Maslov-2003,Lestone-2009}). 
Two widely used current models that contain a statistical description of fission are the code GEMINI$++$ 
\cite{GEMINI-2010} in which fission is described by the transition-state approach of Moretto \cite{Moretto}
and the code SMM (\cite{SMM-2013}, and references therein) in which low-energy fission is described by 
an empirical parametrization and higher-energy fission is treated as one of the possible channels of 
statistical multifragmentation.

Along with the statistical description of fission, the dynamical approach to the fission process
was put forward early on in the seminal paper of Kramers \cite{Kramers-1940}. Based on the assumption
that the deformation degree of freedom can be viewed as a Brownian particle interacting stochastically 
with the single-particle degrees of freedom that constitute a heat bath, Kramers analytically solved 
the appropriate 1D Fokker-Planck equation and predicted that fission is actually retarded relative 
to its rate obtained from purely phase-space arguments.
Along these lines, dynamical approaches based on the dissipative character of the nuclear shape motion
were developed based either on the Fokker-Planck equation or the Langevin equations in more that one 
demensions describing the deformation degrees of freedom. A detailed review of these efforts is given
in \cite{Abe-1996,Frobrich-1998}.
Refined dynamical approaches along these lines have continued with increasing degree of sophistication
and success 
(e.g. \cite{Chaudhuri-2001,Karpov-2001,Nadtochy-2007,Sad-2009,Sad-2011,Nadtochy-2012,Eslam-2014,Ye-2014,
Aritomo-2014a,Aritomo-2014b,Nadtochy-2014,Aritomo-2013,Randrup-2013,Randrup-2011a,Randrup-2011b,Moller-2015z,
Mazurek-2015z}).


Notable to all efforts to descibe fission based on the macroscopic LDM is the prediction of symmetric
mass yields, in contrast to the large body of experimental data  of minor actinides   
that indicate asymmetric mass yield distributions of low-energy fission.
This discrepancy was remedied by Swiatecki \cite{Swiatecki-1955} by the inclusion of a microscopic
correction to the macroscopic LDM part of the potential energy surface (PES) of the nucleus.
A detailed shape-dependent macroscopic-microscopic PES was obtained by Strutinsky 
\cite{Strutinsky-1967,Strutinsky-1968} forming the basis of the successfull
shell-correction approach to the PES.
  
The Strutinsky macroscopic-microscopic approach, due to its simple physical foundation and numerical
flexibility,
has seen continuous development and success. Detailed relevant reviews concerning the description of fission
barriers can be found in \cite{Moller-2009,Pomorski-2013,Moller-2015}.  
The most detailed prediction of PES based on the  macroscopic-microscopic approach was performed by
M\"{o}ller et al. on a five-dimensional deformation space \cite{Moller-2001,Moller-2009}. 
This detailed PES description forms the basis of the recent dynamical description of fission in 
the limit of strong dissipative coupling (Smolutsovski limit) in which the fission process 
resembles a random walk on the multidimentional PES \cite{Randrup-2013,Randrup-2011a,Randrup-2011b,Moller-2015z}. 


Apart from the macroscopic-microscopic description of the heavy-element PES relevant to the 
description of nuclear fission, fully microscopic approaches based on the nuclear 
Density Functional Theory (DFT) have been developed. Some recent representative works
are \cite{Rodriguez-2014,Sad-2014,Sad-2013,Giuliani-2014,Schunck-2014,Schunck-2015, Giuliani-2013,McDonnell-2012,McDonnell-2014} in which
extensive reference to previous works can be found. We also report the recent work \cite{Pei-2014}
on unconstrained DFT calculations in which properties of very deformed nuclei pertinent
to fission are described.


Whereas the static properties of the PES are very well accounted for by the modern DFT-type
approaches, the microscopic description of the full dynamics of the fission process still
remains a daunting project for nuclear theory.

Two main quantal approaches have been adopted in the past to describe the fission dynamics. 
First in 1978, the time-dependent Hartree-Fock (TDHF) theory was applied to the fission 
process \cite{Negele-1978}.
However, as documented  in recent studies of heavy-ion collisions 
\cite{TDHF-HI-2013,TDHF-HI-2011,TDHF-HI-2010}, 
the TDHF approach, being a one-body (mean-field) approach 
has essentially no correlations beyond the mean field and, as such,
cannot fully describe fluctuations 
(as, e.g., encountered in nucleon exchange or fragment formation).
The lack of fluctuations also affects the correct triggering of scission  
in the time evolution of a deforming nucleus.

The second approach is based on the adiabatic approximation and involves the adiabatic TDHF theory
\cite{Baranger-1978} and the time-dependent generator coordinate method (GCM) \cite{Goutte-2005}.
In these approaches, the adiabadic hypothesis is invoked for the fission process, namely,  
that the nucleonic degrees of freedom are fully equilibrated during the slow evolution
of the deforming nucleus over the macroscopic PES determined by the appropriate deformation degrees
of freedom.  We note that the adiabatic approximation is also inherent to the Langevin 
(or Fokker-Plank) type of approaches mentioned earlier.
Under this approximation, the dynamics is described up to the scission configuration. At this point
a sudden approximation is invoked (as in the early scission model \cite{Wilkins-1976}) in order to
obtain the fission fragments and their characteristics (mass, charge, kinetic energy).
However, as stated in recent works \cite{Rizea-2013,Simenel-2014}, ignoring the non-adiabatic effects during
and past scission leads to limitations in the predictive power of the models regarding the fragment
characteristics (e.g., the kinetic energy).   
  
The recent dynamical study \cite{Simenel-2014} attempts to bridge the two regimes.
The adiabatic phase of the fission process is described with a static mean-field (DFT) approach
and the nonadiabatic phase is carefully described with TDHF. It is found that the proper treatment
of the nonadiabatic phase results in an accurate description of the kinetic energy and excitation energy
of the resulting fission fragments.
Before closing the above review on dynamical approaches, we also report efforts to describe 
spontaneous fission employing TDHF and  Feynman's path integral approach \cite{Negele-1989}, as well as
the semiclassical equivalent method employing Vlasov's equation \cite{Bonasera-1997a,Bonasera-1997b}.


The preceeding rather limited overview of the extended literature on nuclear fission dynamics
clearly indicates that a full microscopic description of the fission
process is unusually challenging and, as of today, ``it has not been possible to establish a computationally
feasible framework capable to describe real nuclei with realistic interactions'' \cite{Rodriguez-2014}.  
Nevertheless, we can see that substantial progress in both macroscopic, as well as quantal dynamical
approaches continues vividly.
Motivated by the current situation regarding fission dynamics,  in the present work,
we initiated a study of fission based on the semiclassical microscopic N-body Constrained Molecular Dynamics
(CoMD) model
\cite{Papa-2001}  in regards to its ability to describe the full dynamics of the fission process
in proton-induced reactions on Th and U from low to high energies.

In the following discussion, we classify, as customary, the proton-induced fission reaction 
according to the proton energy E$_{p}$ as: a) low-energy when  E$_{p}$ $<$ 20 MeV, b) intermediate energy when
20 MeV $<$ E$_{p}$ $<$ 200 MeV and c) high energy when E$_{p}$ $>$ 200 MeV.
The present paper has the following structure. In section II, we highlight the basic aspects of 
the CoMD code and present the way that the code is applied to nuclear fission. 
In section III, we present results of fission calculations for proton-induced reactions on : 
a) $^{232}$Th at 27 MeV and 63 MeV, 
b) $^{235}$U at 10 MeV, 30 MeV, 60 MeV and 100 MeV, and 
c) $^{238}$U at 100 MeV and 660 MeV. 
The fission fragment mass yield distributions are presented, as well as total fission cross sections, total energies,
neutron multiplicities  and fission times with respect to incident proton energy.
In section IV, discussion of the results and conclusing remarks are given. 

\section{ Description of the Theoretical Model } 

The theoretical model employed in this work is the microscopic
Constrained Molecular Dynamics (CoMD) model originally designed for reactions near and below
the Fermi energy \cite{Papa-2001,Maru-2002,Papa-2005}. 
Following the general approach of Quantum Molecular Dynamics (QMD) models \cite{QMD-1991}, 
in the CoMD code nucleons are described as localized Gaussian wave packets.
The wave function of the N-body nuclear system is assumed to be the product of these single-particle
 wave functions. With the Gaussian description, the N-body time-dependent Sch$\ddot{o}$dinger equation 
 leads to (classical) Hamilton's equations of  motion for the centroids of the nucleon wavepackets. 
The potential part of the Hamiltonian consists of a Skyrme-like  effective interaction
and a surface term. 
The isoscalar part of the effective  interaction corresponds to a nuclear matter
compressibility of K=200 (soft EOS) or K=380 (stiff EOS).
For the isovector part, several forms of the density dependence of
the nucleon-nucleon symmetry potential are implemented.
Two of them will be used in the present work:  the ``standard'' potential [red (solid) lines] and 
the ``soft'' potential  [blue (dotted) lines] in the figures that follow.
These forms correspond to a dependence of the symmetry potential on
the 1 and the 1/2 power of the density, respectively  (see, also, \cite{Papa-2013} 
and references therein).

We note that in the CoMD model, while not explicitly implementing antisymmetrization 
of the N-body wavefunction, a constraint in the phase space occupation for each nucleon
is imposed, effectively restoring the Pauli principle at each time step of the (classical)
evolution of the system. This constraint restores the fermionic nature 
of the nucleon motion in the evolving nuclear system.
More specifically, at each time step, and for each nucleon, the presense of neighboring nucleons 
is determined in phase space. If the phase space occupation probability is greater than 1, 
then the code changes the direction (not the magnitude) of the nucleon momentum,  so that the 
total momentum and kinetic energy are conserved \cite{Papa-2001}.

The short range (repulsive) nucleon-nucleon interactions are described as individual nucleon-nucleon
collisions governed by the nucleon-nucleon scattering cross section, the available phase space
and the Pauli principle, as usually implemented in transport codes (see, e.g. \cite{Bonasera-1994}).
The present CoMD version fully preserves the total angular momentum (along with linear momentum and energy),
features which are critical for the accurate description of observables from heavy-ion collisions and,
for the present study, the fission dynamics.

The ground state configurations of  the target nuclei were
obtained with a simulated annealing approach and were tested for stability
for long times (2000--3000 fm/c). These configurations were used in the 
subsequent particle-induced fission simulations. 


In the calculations of the present work, the CoMD code was used mainly with its standard parameters. 
The soft density-dependent isoscalar potential was chosen (K=200),  as well as the “standard” and “soft” 
symmetry potentials, as mentioned above. The surface term was set to zero.

For a given p-induced reaction, a total of 3000--5000 events were collected. For each event, 
the impact parameter
of the collision was chosen in the range b = 0--6 fm, following a triangular distribution. 
Each event was followed in time up to 15000 fm/c ( 5.0$\times$10$^{-20}$ sec )
The phase space coordinates were registered every 50 fm/c. 
At each time step, fragments were recognized with the minimum spanning tree method \cite{Papa-2001}, 
and their properties were reported. 
From this information, we obtained information on the evolution of the fissioning system and the properties
of the resulting fission fragments.
We mention that we consider as fission time (t$_{fission}$) the time interval between the implantation of the
proton in the target nucleus and the emergence of the two fission fragments.
We allowed an additional time of 2000 fm/c after scission for the nascent fission fragments to de-excite.
(We varied this time interval from 2000 to 5000 fm/c and we did not notice an appreciable change in the 
characteristics of the fission fragments.)
Thus, in the following discussion, for each event, the fission fragment properties are reported and studied
2000 fm/c after scission. 

A typical time evolution of a fissioning system as predicted by CoMD is presented in Fig. 1.
The figure refers to p-induced fission of $^{232}$Th at 63 MeV and gives
a three dimensional representation of the fissioning system in the center of mass at three time instants. 
At 0 fm/c (Fig. 1a), the proton approaches the target nucleus $^{232}$Th. At 2000 fm/c (Fig. 1b) the nucleus has been substantially deformed. This configuration is near or past the saddle point.
(For this event scission occurs at t$_{fission}$=2500 fm/c.)
At 4000 fm/c (Fig. 1c), we observe the two fission fragments and the emission of two neutrons departing 
nearly perpendicular to the fission axis.

In Fig. 2, the time evolution of the axial quadrupole moment Q$_{20}$ of the fissioning system (Fig. 2a) and its mean radius (Fig. 2b) are presented for the same event. Both quantities increase with time and indicate the course of the system toward fission. (We note that the decreasing value of Q$_{20}$  after scission is attributed to the continuous rotation of the deformed nucleus and the resulting fission fragments.) 

\section{  Results and comparisons }

\subsection{Mass yields: Low and Intermediate Energy} 

We begin our study of the behavior of the CoMD code with comparisons 
to the recent experimental data described in \cite{Demetriou-2010,Isaev-2008}.
First we show the CoMD calculations for the proton  induced fission of $^{232}$Th at energies 
27 MeV and 63 MeV, using the “standard” and the “soft” symmetry potential. 
In Fig. 3, the mass yield distributions  are illustrated for the reaction at 27 MeV. 
In the experimental data (full points), we observe the asymmetric nature of the fission mass yield,
as expected for low-energy fission of minor actinides. 
We compare the data with our calculations (open points with statistical errorbars) with the standard symmetry 
potential (Fig. 3a),
as well as the soft symmetry potential (Fig. 3b). With both selections of the symmetry potential, 
we observe a merely symmetric mass distribution with a rather flat top.
No clean sign of an asymmertic mass yield distribution is seen in the CoMD calculations.
A hint for asymmetric distribution may be implied in Fig. 3b. 
At this point, we note the finding of Nadtochy et al. \cite{Nadtochy-2013} that in dynamical
Langevin calculations of fission, dominance of asymmetric mass splits relative to the 
(expected) symmetric mass split may occur as a result of the dissipative dynamical behavior 
of the system. A similar suggestion that asymmetric fission may result from the hydrodynamical
behavior of the system was first reported in \cite{Nix-1969} before strong shell effects were 
considered responsible for the asymmetric mass yield distribution of actinides 
\cite{Strutinsky-1967}. 
 

The main reason for the symmertic mass yield distribution obtained by the CoMD code is that
the nucleon-nucleon interaction in the model does not include 
spin dependence, thus the resulting mean-field potential does not contain a spin-orbit contribution.
Thus, the model does not predict the correct shell effects in the single-particle motion of the deforming nucleus,
which are necessary to  lead to the asymmetric fission of $^{232}$Th. 
A closer inspection of the two yields calculations, shows that they are slightly different but neither of them
tends to resemble the experimental distribution.

We wish to comment that, while in the present implementation of the CoMD model the interaction has no spin dependence,
(thus CoMD cannot describe the correct shell effects), the code  emulates the quantum behavior of the 
deforming nuclear system, thus we should expect shell effects (at the mean-field level) which would correspond 
to those obtained by a deforming harmonic oscillator or Woods-Saxon potenial (without a spin-orbit term)
\cite{Prussin-book}.
We also mention that application of the CoMD approach to light atoms has successfully reproduced the 
electronic binding energies, as well as electron radii revealing shell structure \cite{Kimura-2005z}. 
A study of shell effects in the present implementation of CoMD applied to nuclei has not been performed to date. 
Such a study with the present CoMD code and a possible extension of it with spin dependence 
(in the spirit of recent work on BUU \cite{BUU-spin-2014}) will be undertaken by us in the near future.


In Fig. 4, the mass yield distribution for the same reaction at proton energy 63 MeV is presented. 
It is evident that the structure of the experimental mass yield curve tends to become more symmetric 
at this higher energy. 
This is to be expected, because
as the proton energy and, thus, the excitation energy of the fissioning system increases, 
shell effects will begin to fade (see e.g. \cite{Randrup-2013,Chaudhuri-2015z}).  
However, it seems that this beam energy is not high enough to completely wash out the shell effects,
as two asymmertic shoulders appear in the experimental mass yield curve. 
In Fig. 4a, we show the CoMD calculations with the standard symmetry potential and in Fig. 4b,  
the soft symmetry potential.  As in the lower energy case (Fig. 3), the two choices of the symmetry 
potential  do not lead to  substantial differences on the mass yield shape. 

We note that at this higher energy, as the asymmetric mass split is  attenuated
and the symmetric contribution is enhanced, an overall improvement in the agreement between 
our CoMD calculations and the experimental data is obtained.
Thus, with the current implementation of the CoMD code, intermediate-energy fission mass yields
may be correctly descibed.  We will explore this behavior with other fissioning systems 
in the following.


We continue our comparisons with the recent work \cite{Mulgin-2009} on the proton-induced fission of  $^{235}$U 
at proton beam energies of 10 and 30  MeV. 
In Fig. 5, we present the calculated mass yield distributions (open points) of the proton induced fission of
$^{235}$U at 10 MeV, using the two forms of the symmetry potential, the standard (Fig. 5a) and the soft (Fig. 5b). 
The experimental data of \cite{Mulgin-2009} are presented by closed points. 
We note that the yield data in \cite{Mulgin-2009} are in arbitrary units, so we multiplied the yields
with a factor of 7,  for both reactions to make them comparable with our calculated cross section results.
In the experimental data, the prevalence of the asymmetric fission mode is obvious for 
this low-energy fission reaction.
Our calculations with the standard symmetry potential indicate a rather symmetric distribution (Fig. 5a).
Interestingly, the calculations with the soft symmetry potential indicate a hint of asymmetric fission 
(Fig. 5b).
However, as already discussed above, in the absence of the correct shell effects from the CoMD potential, 
we do not expect to obtain the correct asymmetric mass distribution.
In the comparisons of Fig. 5, we observe that the CoMD calculations result in a wider mass yield
distribution as compared with the data. Apart from a possible calculational aspect, this may also 
point to a limitation in the data toward asymmetric mass splits.
 
  
In Fig. 6,  we present the mass yield distribution from the p-induced fission of $^{235}$U 
at proton energy 30 MeV. The experimental data come from \cite{Mulgin-2009}. 
In the CoMD calculations,  we again used the two forms of the symmetry potential,
the standard (Fig. 6a) and the soft (Fig. 6b). 
Conclusions similar to those from Fig. 5 can be drawn. 
However, going from 10 MeV to 30 MeV proton energy, in the experimental mass yield distributions 
we see that the peak-to-valley ratio is reduced as the symmetric contribution  increases relative to the 
asymmetric contribution. 
Thus, in Fig. 6, the CoMD calculations (yielding a symmetric mass distribution) are obviously in better 
agreement with the experimental data.

Furthermore we explored the behavior of CoMD at higher energy for the same system.
In  Fig. 7 we show the CoMD calculations for proton induced fission of $^{235}$U at 60 MeV,
again with the two forms of the symmetry potential, standard (Fig. 7a) and soft (Fig. 7b).
We compared our calculations with available experimental data of proton induced fission 
of $^{238}$U (not of $^{235}$U) at this beam energy taken from \cite{Dui-2001} 
(and normalized to our calculated  cross sections, as in Figs. 5 and 6). 
We note that the small difference in the number of neutrons of the fissioning systems at this higher
energy is not expected to substantially affect the mass yield comparisons of Fig. 7.
The experimental distribution for U indicates that the asymmetric fission mode still prevails.
This yield curve presents a plateau, in contrast to the Th distribution at comparable energy
that has a symmetric peak and two shoulders at asymmetric mass splits (Fig. 4).
Our calculations, as expected, indicate a symmetric peak, as we saw in the lower energy cases 
(Fig. 5, 6) of the proton induced U fission.
 

\subsection{Mass yields: High Energy} 

We continue our investigations with the application of CoMD to high energy fission reactions 
with protons.  
We remind that the CoMD code has been successfully applied to the description of 
a large variety of nuclear reactions
(e.g., \cite{Papa-2001,Amorini-2009,GS-CoMD-2010,GS-RIB-2014,GS-NZ-2014}).
As a fully dynamical code, we expect that it may perform well also with spallation-type
reactions with high-energy protons. 

Toward this direction, we performed calculations for the proton induced fission of $^{238}$U 
at 660 MeV proton energy. 
The motivation comes from the importance of this energy range in ADS-type applications 
\cite{Aliberti-2004,Koning-2005} and,  specifically,  the recent experimental data for this reaction 
reported in \cite{Karapetyan-2009,Balabekyan-2010,Deppman-2013a,Deppman-2013b} obtained 
by off-line gamma-ray techniques.
In Fig. 8, we show the experimental data (full symbols) that despite the 
large experimental fluctuations indicate a prevailing symmetric 
fission mode. (We point out that the yield axis is logarithmic in this figure).
Our CoMD calculations, with the two forms of the symmetry potential (Fig. 8a standard, Fig. 8b soft)
are in reasonable  agreement with the experimental data near symmetric mass splits.

We note that the shape of the experimental mass distribution is characterized by two low-yield very asymmetric
fission components ( ``super-asymmetric'' fission \cite{Deppman-2013b}). 

Our calculations show an overall symmetric curve that is wide enough to contain these superasymmetric 
mass splits, predicted, however, with larger cross sections than the data. 
Our calculations resemble the wide symmetric mass yields observed resently in high-quality
mass spectrometric data obtained in inverse kinematics at relativistic energies
\cite{GSI-data-2006,GSI-data-2013,Spallation-2009}. 
Comparing the data of \cite{Karapetyan-2009,Balabekyan-2010} 
with the higher-energy inverse-kinematics data (e.g. \cite{GSI-data-2006}), 
we speculate that the former data may be incomplete 
due to the nature of the measurements and we suggest that measurements of this very important reaction
at $\sim$600 MeV be performed in inverse kinematics in the same fashion as the higher-energy data.
 

We wish to point out that the lower-energy fission data discussed above (Figs. 3-7) have  been 
acquired with standard fission on-line counters,  which cannot provide information on the atomic number Z of 
the fission fragments. 
We remind that to obtain Z information, either mass spectrometric tecniques (mainly in inverse kinematics)
or off-line gamma-ray methods have to be used.
As already mentioned, the data of \cite{Karapetyan-2009,Balabekyan-2010} were
obtained with gamma-ray techniques and thus can provide information on the Z-A correlation
of the observed fission fragments.

In Fig. 9, we first present the experimentally observed  mean Z  (Fig. 9a) 
as well as the standard deviation of the Z distribution (Fig. 9b.)
with respect to the mass number A of the fission fragments.  
Our CoMD calculations [solid (red) line] show that the fission fragments are more neutron-rich relative
to the experimental data. 
Due to the fact that, for this high-energy reaction, the  code causes one or two protons to be emitted before the scission point, we made a selection concerning the charge of the fissioning nucleus (Z=93) so that it corresponds to no pre-scission
proton emission at the time of scission. With this selection, the CoMD calculations [dashed (blue) line] are in better agreement with the data, especially for the heavier fragments.
From the above comparison for this high-energy reaction, we may conclude that the fission fragments, as obtained
2000 fm/c after scission, may still contain enough excitation energy to further evaporate (predominantly) neutrons
and, thus, move closer to the data in the Z--A plot.
An explicit de-excitation of these fragments with a standard de-exciation code (e.g. \cite{GEMINI-2010,SMM-2013})
was not performed in the present exploratory work,  
but will be performed by us in the near future within our plan of detailed studies of high-energy p-induced fission
of actinides.
In Fig. 9b, our calculations of the standard deviation are higher in comparison with the experimental data. 
However, when the selection of the fissioning nucleus is made, so that it corresponds to no pre-scission
proton emission (Z=93),  the calculations are in better agreement with the data, 
despite the large fluctuations due to the limited statistics of the calculations after the 
imposed selection.


\subsection{Fission Cross Sections}

After the presentation of the mass yield distribution, which is one of the most characteristic 
observables of fission reactions, we continue our investigation with several other fission
observables, starting with the total fission cross sections. 
In Fig. 10, calculated total fission cross sections are presented as follows.
The full (red) triangles connected with a full line refer to the p-induced fission of $^{232}$Th 
at the two energies 27 and 63 MeV.
The full (red) circles connected with a full line represent the p-induced fission of
$^{235}$U at 4 energies  (10, 30, 60 and 100 MeV). 
The full (red) squares connected with the full line show the p-induced fission of $^{238}$U 
at two proton energies (100 and 660 MeV). 
The CoMD calculations correspond to the “standard” symmetry potential.
We note that the points at E$_{p}$=660 MeV are displayed at E$_{p}$=160 MeV in Figs. 10--15,
for convenience of presentation.

The available experimental data for $^{232}$Th are shown with closed (black)  triangles connected 
with dotted lines.
The experimental point for $^{235}$U at E$_{p}$=60 MeV is shown with a closed (black) circle
and for $^{238}$U at E$_{p}$=660 MeV with closed (black) square. 

Concerning the fission of thorium, we observe that increasing the proton energy (and thus the excitation 
energy of the fissioning nucleus), there is only a slight increase in the calculated cross section. 
However, the experimental data show an increase of approximately 30$\%$, which our calculations do not 
reproduce. 
For the fission of $^{235}$U, in the calculations we observe a jumb of the cross section, 
with increasing proton energy from 30 MeV to 60 MeV. The experimental  point at 60 MeV 
is 20\% larger than our calculated point. 
At higher  energies, for the proton induced fission of $^{238}$U , the total fission cross section is 
rather constant and in rough agreement with the experimenal data within errorbars.  


In Fig. 11, we present the ratio of the fission cross section to the heavy-residue cross section
as a function of the proton energy. This ratio is a very sensitive observable for the relative
importance of fission as a deexcitation path for the nuclei examined.

The CoMD calculations (with the standard symmetry potential) are shown with the closed (red) symbols
connected with solid (red) lines with exactly the same correspondence as in Fig. 10.
In addition, CoMD calculations with the soft symmetry potential are also shown in this figure 
with closed (blue) symbols connected with dotted (blue) lines (such calculations were not shown in 
Fig. 10 because they would nearly overlap with the ones shown). 
 
For the fission of thorium we observe an increase in the  ratio from 27 to 63 MeV and a 
rather weak sensitivity to the choice of the symmetry potential, the ratio being slightly
larger at the energy of 63 MeV with the choice of the soft symmetry potential.
For the fission of $^{235}$U, the ratio increases from 10 to 30 MeV and then 
diminishes at the higher energies of 60 and 100 MeV. 
Decerasing trend also exhibits the  ratio for $^{238}$U from 100 to 660 MeV, possibly 
pointing to the increasing role of fast evaporation processes for the more excited nuclei
involved in the higher energy reactions. 

Finally, focusing our attention to the behavior of the calculated ratio with the soft symmetry
potential for the  $^{235}$U and  $^{238}$U isotopes, interestingly we observe that this ratio
is substantially larger than the corresponding ratio calculated with the standard symmetry
potential. Further detailed investigation of the  features of this bevavior is in line.
However, from the present work we conclude that the ratio is a rather sensitive observable
of the density dependence of the nucleon-nucleon symmetry potential, and thus, of the nuclear 
symmetry energy, which is a topic of current importance 
in regards to studies of the nuclear equation of state (e.g. \cite{Bonasera-2014,Kohley-2014,Horowitz-2014}).


\subsection{Total Fission Kinetic Energy}


In the following we will examine the mean total kinetic energy of the fission fragments 
as a function of the proton energy for the studied reactions.  This is an important kinematical
observable characterizing on average the degree of deformation, the compactness and the asymmetry at scission 
of the fissioning  system  and offers an important testing ground of the     
overall dynamical description offered by the employed code. 
In Fig. 12,  we illustrate the mean total kinetic energy of the fission fragments for the  aforementioned
fission reactions. The symbols correspond to the same reactions as in Figs. 10 and 11.

For the p-induced fission of $^{232}$Th, the calculated kinetic energy [closed (red) triangles] 
is nearly the same for the two studied reactions at 27 and 63 MeV. 
The values are lower than the experimental data [closed (black) triangles] which
indicate an increase of the kinetic energy with the increase of the excitation energy of the fissioning system. 
For the p-induced fission of U, the calculated kinetic energy [closed (red) circles] increases slightly when 
we go from 10 to 30 MeV.
At higher energies,  there is a small but continuous decreasing trend. 
The available experimental data \cite{Mulgin-2009,Isaev-2008} for the first three proton energies
[closed (black) circles] are higher  than our calculations, 
with an increasing trend from 30 to 60 MeV.
Along with the proton-induced reactions studied in this work, we mention the recent experimental
data of \cite{Loveland-2014} on the fission kinetic energies of the neutron-induced fission of
$^{235}$U in a broad energy range below 50 MeV. We note that the measured fission energies in
the neutron energy range 30--45 MeV are approximately 162 MeV, in overall agreement (albeit higher)
with our calculation for the p-induced $^{235}$U fission (Fig. 12).

We can relate the lower kinetic energy obtained by CoMD, as compared to the experimental data,
to the observation that the CoMD code implies emission of about two (on average) pre-scission
protons  even at the lower energy fission reactions at 10 and 30 MeV. 
This is unphysical as can be concluded from our up-to-date experimental and theoretical 
understanding of the low-energy fission process. We discuss this feature quantitatively 
in the following.
Furthermore, in all reactions studied, the CoMD calculations of the total energy with the soft symmetry potential
[(blue) points connected with (blue) dotted lines] are lower than the corresponding ones with the 
standard symmetry potential [(red) points connected with (red) solid lines].
This may point to a scission configuration with a more elongated shape (and longer neck) in the soft case,
as compared to a more compact shape (and a shorter neck) in the standard case.   
\subsection{Pre-scission and Post-scission Particle Emission}


In Fig. 13 we show the calculated average pre-scission, post-scission and total proton  
multiplicity from the studied reactions.
Of course,  the pre-scission proton emission should not be present at the lower energies,
but it should compete with the pre-scission neutron emission at higher energies. 
A clear increasing trend is present in the calculations with a substantial increrase at 
the highest energy of 660 MeV.  
We think that further detailed investigation is necessary to understand this 
feature of pre-scission proton emission of the code at low energies (see also below the 
corresponding situation for neutrons).


We now discuss the predictions of the CoMD code concerning the pre-scission, post-scission
and total neutron multiplicity.
We remind that the pre-scission neutron multiplicity serves as a clock 
of the evolution of the fissioning system up to the moment of scission, whereas the 
post-scission neutron multiplicity can be directly related to the excitation energy of 
the nascent fission fragments \cite{Hinde-1992}.  
Both quantities can be obtained experimentally with proper, albeit especially difficult, measurements 
and model analysis and can offer very sensitive observables for any dynamical model of fission.

From the present study with the CoMD code, we show in Fig. 14   the pre-scission, 
post-scission and total neutron multiplicities versus proton energy. 
These quantities show an overall increasing trend with increasing  energy 
for the studied fissioning systems. 

More specifically, for the p-induced fission of $^{232}$Th, the calculated pre-scission neutron 
multiplicity [closed (red) triangles] is nearly 3 at both energies 27 and 63 MeV (showing a small increassing trend).
The experimental value is nearly 1 at the energy of 27 MeV, and thus about two units lower
than the calculation. 
This observation now reveals that the CoMD calculation predicts a larger 
pre-scission neutron multiplicity (by about two units),  as was the case for the pre-scission 
proton multiplicity that we examined before (that was assumed responsible for the observed 
lower total kinetic energy of the fission fragments, as compared to the experimental values).
This conclusion, along with the corresponding one in regards to pre-scission  proton emission, 
calls for further detailed study of  the parameters of the CoMD code.  We speculate that a 
careful fine-tuning of the surface term may be necessary to suppress the observed unrealistic
feature of both pre-scission proton and neutron emission of the code at low energies.

Furthermore, we note that the experimental value for the pre-scission neutron multiplicity at 
63 MeV is in reasonable agreement with the CoMD calculation.
In regards to the post-scission multiplicities, we can say that the calculation is in fair
agreement with the available data, albeit larger at the higher energy.

For the p-induced fission of $^{235}$U, the calculated pre-scission and post-scission neutron  
multiplicities (Figs. 14a and 14b, respectively) increase steadily as the energy increases from 10 to 100 MeV.
Agreement is seen with the experimental point at 63 MeV for $^{238}$U taken from \cite{Demetriou-2010}.
For the p-induced fission of $^{238}$U at 100 MeV, the pre-scission and post-scission neutron 
multiplicities are slighly higher than the corresponding values for $^{235}$U, of course reflecting 
the larger  neutron content of the former nucleus.
Similar observations pertain to the  total neutron  multiplicities (Fig. 14c) for the  reactions studied.
We also note the agreement of the calculated value with the experimental point at 660 MeV \cite{Deppman-2013a}.
	
\subsection{Fission Timescale}

We will complete the presentation of the predicted fission characteristics 
with a discussion of the fission time as obtained directly by our 
fully dynamical CoMD calculations.
We point out that it is a difficult task to extract the fission time scale from experimental data 
(e.g. \cite{Hinde-1992,Strecker-1990,fission-time}). Furthermore, the exctracted values are  unavoidably model
and method dependent. 
On the other hand, a fully dynamical code, either a macroscopic one, as the current advanced 
Langevin codes (e.g. \cite{Nadtochy-2014,Aritomo-2013,Randrup-2013}) or a microscopic TDHF-type 
code (e.g. \cite{Simenel-2014})
can in principle provide realistic information on the fission time scale,
as long as, the code has been extensively benchmarked by comparison of its predicted fission 
observables  with available experimental data on mass yield distributions, kinetic energies and
neutron multiplicities.
 
In the present study, for the first time the semiclassical N-body CoMD code was tested 
with p-induced fission reactions and, as our presentation so far has indicated, it performed 
in an overall satisfactory manner, especially for intermediate and high energy fission reactions.
The CoMD code, as a fully dynamical code,  naturally describes the  complete dynamical
path of the fission process. Therefore, with CoMD we can determine the fission time in a direct way.

In Fig. 15, we show the extracted average fission time  versus the proton energy of the reactions
studied. 
The presentation of the calculations follows the same pattern as in Figs. 10-12.
Two main groups of points are shown in Fig. 15. The upper group (closed points) corresponds to the 
CoMD predictions of the fission time using the full ensemble of the fissioning nuclei
for each fission reaction.
The lower group (open points) corresponds to the CoMD predictions with a selection made on 
the fissioning nucleus at the moment of fission to have exactly the initial Z value
(thus, assuming no pre-scission proton emission).
Within each group, the points connected with full (red) lines correspond to CoMD calculations
with the standard symmetry potential, whereas the points connected with dotted (blue) lines
are with the soft symmetry potential.
When the selection of no pre-scission proton emission is made (lower group), the chosen
fissioning nuclei are both more excited and more fissile  (compared to those resulting 
after the emission of pre-scission protons), and thus their fission time is 
correspondingly lower.
As a general observation, within both groups, for each reaction, there is an overall 
decreasing trend with increasing proton energy. 
Furthermore, we notice that the choice of the soft symmetry potential results in faster fission 
dynamics. This can be understood qualitatively by the higher potential energy that the soft symmetry 
potential implies for the neutron-rich low-density neck region for a highly deformed fissioning nucleus
and, in turn, can be related to the corresponding lower total fission energy observed in Fig. 12.

For the Th reactions, the fission time is slighly higher than the U reactions
at nearly the same energy, reflecting the lower fissility of the $^{232}$Th nucleus.
We notice that the fission time for the p-induced fission of $^{235}$U is on average longer 
than that of $^{238}$U  at the energy of 100 MeV. 
From Fig. 13 we see that the average pre-scission and post-scission proton multiplicity is similar
for these two reactions,  whereas from Fig. 14 the pre-scission and post-scission neutron  multiplicity
is larger in the latter case. 
We cannot provide a simple explanation of this fission time difference
(which is similar in both groups and, within each group, for each choice of the symmetry potential). 
This difference should be related to the details of the dynamical evolution of the two different
fissioning systems. 

The above discussion indicates that the CoMD code can provide detailed information 
of the fission time under various conditions of excitation energy and fissility.
From our study so far, if we exclude the low energy fission, we expect that from 60 MeV 
and above, the time scale information may be consider realistic. 
It would be very interesting if the present predictions can be compared to experimental 
information when such information becomes available.
Furthermore, it is exciting to notice the sensitivity of the fission time scale to the density 
dependence of  the nucleon-nucleon symmetry potential, and thus isospin part of the nuclear 
equation of state, suggesting this observable as an additional probe of the latter
\cite{Bonasera-2014}. 

\subsection{CoMD Energies of Fissioning Nuclei}

After the above detailed discussion of CoMD observables, we will close our
presentation by examining the evolution of the interaction and kinetic energies
of fissioning nuclei in their course toward scission.

As a representative example, we show in Fig. 16 the results for  
the reaction p(30MeV)+$^{235}$U for the standard symmetry potential
[(red) points connected with solid line] and the soft symmetry potential
[(blue) points connected with dotted line]. 
In order to examine the evolution of the average energies in the course to scission,
and given the broad distribution of fission times (as seen in Fig. 15), in Fig. 16 
the time is referenced with respect to the moment of scission, taken to be at t=0 fm/c.

In Fig. 16a, the event-average nuclear interaction energy is presented, taken to be the sum of
the two-body, three-body, surface and symmetry energy terms. An overall increase 
of the interaction energy is observed as the fissioning system approaches the 
moment of scission, for both the standard and the soft symmetry potentials.
Interestingly, the calculation with the soft symmetry potential results in higher
interaction energy of the fissioning system,  as can be understood by the overall 
larger nuclear symmetry energy in the low-density neck region, and  the overall
more repulsive dymanics implied.
In Fig. 16b, the total Coulomb energy is presented, being nearly the same for the 
two choices of the symmetry potential (only slightly lower for the soft symmetry 
potential, since as we discussed, it results in more elongated configurations).
A monotonic decrease of the Coulomb energy is observed as the fissioning system 
evolves toward scission, this decrease being essentially the main driving force
of the nuclear system to fission.

In Fig. 16c, the total potential energy of the fissioning system, namely the sum of
the nuclear interaction energy and the Coulomb energy, is shown. In Fig. 16d, the kinetic 
energy of the fissioning system is shown, being lower for the soft symmetry potential,
that also results in higher potential energy (Figs. 16a, 16c) involving more elongated 
configurations and more repulsive dynamics  in the low-density neck region.
Finally, in Fig. 16e, the total energy of the fissioning system is shown, being slightly 
higher for the soft symmetry potential. The decrease with time is due to the prescission
emission of neutrons and protons (Figs. 13 and 14).

In Fig. 17, we show the variation of the energies of the fissioning system p+$^{235}$U
with respect  to the change of the incident proton energy from 
10 MeV [(blue)   points connected with dotted line] to
30 MeV [(red)    points connected with solid line] to
60 MeV [(green)  points connected with dashed line]. 
The calculations are with the standard symmetry potential.
In Fig. 17a, we observe that with proton energies of 10 and 30 MeV, 
the average nuclear interaction energy of the fissioning systems 
is nearly similar. The Coulomb energy (Fig. 17c) and the kinetic energy (Fig. 17d)
increase in going from 10 to 30 MeV proton energy. 
It appears that the additional proton energy brought in the fissioning system is stored
as kinetic energy (Fermi motion) and Coulomb energy. This increase is reflected in the
total energy (Fig. 17e).

Interestingly, in going from 30 to 60 MeV proton energy, 
we observe that the nuclear interaction energy is increased by nearly this amount
of energy (about 30 MeV), whereas the Coulomb energy is nearly the same. 
The kinetic energy decreases slighly, which is rather counterintuitive: we would expect 
an increase in the kinetic energy as we saw in going from 10 to 30 MeV proton energy.
Thus, for the case of 60 MeV protons, the additional amount of energy 
brought in by the proton is stored as nuclear potential energy, indicating that essentially
above the Fermi energy, the nuclear (mean-field) potential energy is effectively  momentum depend.
The origin of this momentum dependence is in the Pauli correlations imposed by the CoMD  procedure, 
i.e. the phase-space constraint imposed by CoMD to ensure the Fermionic behavior of
the classically evolving system of Gaussian wave packets (see Section II).
The total energy for the case of 60 MeV protons (Fig. 17e) is consistently above that of the
previous two energies and has a diminishing behavior with time toward scission due to the 
emission of prescission particles, as in the other two energies.
 
The above examination of the CoMD energy variations of the fissioning system provides a 
good check of the consistency and accuracy of the code,  as applied to the description of 
a deforming system as it evolves toward scission.
We wish to point out that from the present calculations we cannot obtain information 
regarding the fission barriers of the involved fissioning nuclei. The reason is that the 
calculations are performed at high enough energy, so that the fissioning systems are above 
the fission barrier expected to be near 6--8 MeV.
In order to obtain average fission barriers with CoMD,  a different methodology has to be followed:
the total energy of a fissioning nucleus has to be obtained as a function of deformation, 
placing the nucleus into a deformed harmonic potential. This  interesting project, however, requires
further computational effort beyond the scope of the present paper that we plan  to undertake 
in the near future. 


\section{Discussion and Conclusions}  

In the present work we employed the semi-classical microscopic N-body code CoMD to
describe   
proton-induced fission of  $^{232}$Th, $^{235}$U and $^{238}$U 
nuclei at various energies.
In retrospect, we chose these nuclei because of the availability of recent literature data and 
because of their significance in current applications of fission. 
We found that the CoMD code in its present implementation is able to describe fission 
at higher energies (i.e., above E$_{p}$=60 MeV) where shell effects are mostly washed out. 
We remind that the effective nucleon-nucleon interaction employed in the code has 
no spin dependence, and thus the resulting mean field has no spin-orbit contribution. 

The total fission cross sections of the studied reactions were rather well reproduced.
Furthermore, the ratio of fission cross sections over residue cross sections
showed sensitivity to the choice  of the nucleon-nucleon symmetry potential
and, thus, to the density dependence of the nuclear symmetry energy. 
Consequently, this ratio can be used as a probe of the nuclear equation of state at low density 
and moderate excitation energy, corresponding to intermediate and high energy light-particle 
induced fission.
Concerning  total fission energies and neutron multiplicities, we found that they were 
rather adequately reproduced by the CoMD code (except at the lower energies).
Finally, information on the fission time scale can be obtained from the present calculations.
The obtained fission times  show a dependence on the excitation energy of the nucleus,
as well as on the choice of the symmetry potential. Thus, the fission time offers one more
observable sentitive to the isospin part of the nucleon-nucleon effective interaction.

In regards to the N/Z properties of the fission fragments, the CoMD code appears to perform 
well for the case of p (660 MeV) + $^{238}$U that we tested in this work.
We note that the majority of the data of p-induced fission reactions have been performed 
in direct kinematics with on-line techniques and, thus, Z information of the fission fragments 
cannot be obtained. 
Apart from off-line gamma-ray techniques, such information has been obtained in studies 
in inverse kinematics mostly at high energies (e.g. \cite{GSI-data-2006}).
In parallel to these experimental efforts, we mention the novel mass-spectrometric
study of the reaction $^{238}$U (6.5 MeV/nucleon) + $^{12}$C in which proton-pickup and other
channels leading to fission were chosen by kinematical reconstruction \cite{Farget-2013,Farget-2014}. 
From an application point of view, it would be very important to obtain experimental information 
in inverse kinematics at proton energies from 100 MeV to 1000 MeV. 
As we mentioned earlier, we plan to perform detailed calculations of high-energy proton-induced fission 
in the near future.
Of course the CoMD code can be used for neutron and other light particle incuced fission reactions, 
heavy-ion fusion/fission reactions, as well as multinucleon-transfer/fission reactions 
and its predictions can be compared with existing or future experimental data. 
We also plan to undertake calculational and experimental efforts in this broad direction
in the future.


To conclude, in the present study, the semiclassical N-body code CoMD was tested for the first time 
with p-induced fission reactions. 
We found that the code performs  in an overall satisfactory manner, providing a description
of the full dynamics of the fission process, especially for intermediate and high energy 
fission reactions.
We suggest that inclusion of spin dependence in the nucleon-nucleon effective interaction 
and further improvements of the code should be implemented toward achieving a satisfactory 
description of lower energy fission in which shell effects play a dominant role. 
We point out that the code parameters, as specified predominantly by ground-state properties of nuclei
and nuclear matter, do not dependent on the specific reactions being explored and, as such, 
the CoMD code can offer valuable predictive power for the dynamics of the fission process  in a broad range 
of excitation energy. 
Consequently, the CoMD code can be used for the study of fission of not only stable nuclei, but also of 
very neutron-rich (or very neutron-deficient) nuclei  which have not been studied experimentally to date
and may provide guidance to upcoming RIB experiments.
Moreover, this possibility can be further exploited in studies of fission recycling 
\cite{Goriely-2013,Panov-2013,Erler-2012},
namely, the upper end of the r-process nucleosynthesis by the fission of the resulting very neutron-rich
heavy nuclei.

\section{Ackowledgements}
\par
 

We are thankful to  M. Papa for his version of the CoMD code,
and to Hua Zheng for his rewritten version of the CoMD.
We are also thankful to W. Loveland for his enlighting comments
and suggestions on this work.
Furthermore, we wish to acknowledge the motivation and recent discussions
with Y.K. Kwon and K. Tshoo of the KOBRA team of RISP. 
Financial support for this work was provided, in part, by
ELKE Research Account No 70/4/11395 of the National and Kapodistrian 
University of Athens. 
M.V. was 
supported by the Slovak Scientific Grant Agency under contracts 2/0105/11
and 2/0121/14 and by the Slovak Research and Development Agency under contract
APVV-0177-11.


\bibliography{fission_study}                       

\begin{thebibliography}{118}
\expandafter\ifx\csname natexlab\endcsname\relax\def\natexlab#1{#1}\fi
\expandafter\ifx\csname bibnamefont\endcsname\relax
  \def\bibnamefont#1{#1}\fi
\expandafter\ifx\csname bibfnamefont\endcsname\relax
  \def\bibfnamefont#1{#1}\fi
\expandafter\ifx\csname citenamefont\endcsname\relax
  \def\citenamefont#1{#1}\fi
\expandafter\ifx\csname url\endcsname\relax
  \def\url#1{\texttt{#1}}\fi
\expandafter\ifx\csname urlprefix\endcsname\relax\def\urlprefix{URL }\fi
\providecommand{\bibinfo}[2]{#2}
\providecommand{\eprint}[2][]{\url{#2}}

\bibitem[{\citenamefont{Murray and King}(2012)}]{oiltip}
\bibinfo{author}{\bibfnamefont{J.}~\bibnamefont{Murray}} \bibnamefont{and}
  \bibinfo{author}{\bibfnamefont{D.}~\bibnamefont{King}},
  \bibinfo{journal}{Nature} \textbf{\bibinfo{volume}{481}},
  \bibinfo{pages}{433} (\bibinfo{year}{2012}).

\bibitem[{\citenamefont{Bouchard}(2007)}]{Bouchard-2007}
\bibinfo{editor}{\bibfnamefont{J.}~\bibnamefont{Bouchard}}, ed.,
  \emph{\bibinfo{title}{International Conference on Nuclear Data for Science
  and Tecnology, 2007, DOI:10.1051/ndata:07718}} (\bibinfo{year}{2007}).

\bibitem[{NEA()}]{NEA-2008}
\bibinfo{note}{M. Salvatores, NEA Report No. NEA/WPEC-26, 2008 (unpublished)}.

\bibitem[{\citenamefont{Aliberti et~al.}(2004)\citenamefont{Aliberti,
  Palmiotti, Salvatores, and Stenberg}}]{Aliberti-2004}
\bibinfo{author}{\bibfnamefont{G.}~\bibnamefont{Aliberti}},
  \bibinfo{author}{\bibfnamefont{G.}~\bibnamefont{Palmiotti}},
  \bibinfo{author}{\bibfnamefont{M.}~\bibnamefont{Salvatores}},
  \bibnamefont{and} \bibinfo{author}{\bibfnamefont{C.~G.}
  \bibnamefont{Stenberg}}, \bibinfo{journal}{Nucl. Sci. Eng.}
  \textbf{\bibinfo{volume}{146}}, \bibinfo{pages}{13} (\bibinfo{year}{2004}).

\bibitem[{\citenamefont{Koning et~al.}(2004)\citenamefont{Koning, Hilaire, and
  Duijvestin}}]{Koning-2005}
\bibinfo{editor}{\bibfnamefont{A.~J.} \bibnamefont{Koning}},
  \bibinfo{editor}{\bibfnamefont{S.}~\bibnamefont{Hilaire}}, \bibnamefont{and}
  \bibinfo{editor}{\bibfnamefont{M.~C.} \bibnamefont{Duijvestin}}, eds.,
  \emph{\bibinfo{title}{Proceedings of the International Conference on Nuclear
  Data for Science and Tecnology, Santa Fe, New Mexico, 2004, edited by R.
  Haight et al. AIP Conf. Proc. No. 769 (New York, USA 2005), pp. 1154-1159}}
  (\bibinfo{year}{2004}).

\bibitem[{med()}]{medical-2014}
\bibinfo{note}{Nuclear Physics for Medicine -- NuPECC, accessible at:
  www.nupecc.org/pub/npmed2014.pdf}.

\bibitem[{\citenamefont{Blumenfeld et~al.}(2013)\citenamefont{Blumenfeld,
  Nilsson, and Duppen}}]{RIBreview-2013}
\bibinfo{author}{\bibfnamefont{Y.}~\bibnamefont{Blumenfeld}},
  \bibinfo{author}{\bibfnamefont{T.}~\bibnamefont{Nilsson}}, \bibnamefont{and}
  \bibinfo{author}{\bibfnamefont{P.~V.} \bibnamefont{Duppen}},
  \bibinfo{journal}{Phys. Scr. T} \textbf{\bibinfo{volume}{152}},
  \bibinfo{pages}{014023} (\bibinfo{year}{2013}).

\bibitem[{ATL()}]{ATLAS}
\bibinfo{note}{ATLAS main page: \\ www.phy.anl.gov/atlas/facility/index.html}.

\bibitem[{EUR()}]{EURISOL}
\bibinfo{note}{EURISOL main page: www.eurisol.org}.

\bibitem[{\citenamefont{Schmidt et~al.}(2007)\citenamefont{Schmidt, Keli\'{c},
  Luki\'{c}, Ricciardi, and Veselsky}}]{EURISOL-2007}
\bibinfo{author}{\bibfnamefont{K.}~\bibnamefont{Schmidt}},
  \bibinfo{author}{\bibfnamefont{A.}~\bibnamefont{Keli\'{c}}},
  \bibinfo{author}{\bibfnamefont{S.}~\bibnamefont{Luki\'{c}}},
  \bibinfo{author}{\bibfnamefont{M.}~\bibnamefont{Ricciardi}},
  \bibnamefont{and} \bibinfo{author}{\bibfnamefont{M.}~\bibnamefont{Veselsky}},
  \bibinfo{journal}{Phys. Rev. ST} \textbf{\bibinfo{volume}{10}},
  \bibinfo{pages}{014701} (\bibinfo{year}{2007}).

\bibitem[{RIS()}]{RISP}
\bibinfo{note}{RISP main page: www.risp.re.kr/eng/pMainPage.do}.

\bibitem[{\citenamefont{Tshoo et~al.}(2013)\citenamefont{Tshoo, Kim, Kwon
  et~al.}}]{RISP-2013}
\bibinfo{author}{\bibfnamefont{K.}~\bibnamefont{Tshoo}},
  \bibinfo{author}{\bibfnamefont{Y.~K.} \bibnamefont{Kim}},
  \bibinfo{author}{\bibfnamefont{Y.~K.} \bibnamefont{Kwon}},
  \bibnamefont{et~al.}, \bibinfo{journal}{Nucl. Instrum. Methods Phys. Res. B}
  \textbf{\bibinfo{volume}{317}}, \bibinfo{pages}{242} (\bibinfo{year}{2013}).

\bibitem[{\citenamefont{Goriely}(2015)}]{Goriely-2015}
\bibinfo{author}{\bibfnamefont{S.}~\bibnamefont{Goriely}},
  \bibinfo{journal}{Eur. Phys. J. A} \textbf{\bibinfo{volume}{51}}
  (\bibinfo{year}{2015}).

\bibitem[{\citenamefont{Goriely et~al.}(2013)}]{Goriely-2013}
\bibinfo{author}{\bibfnamefont{S.}~\bibnamefont{Goriely}} \bibnamefont{et~al.},
  \bibinfo{journal}{Phys. Rev. Lett.} \textbf{\bibinfo{volume}{111}},
  \bibinfo{pages}{242502} (\bibinfo{year}{2013}).

\bibitem[{\citenamefont{Panov et~al.}(2013)\citenamefont{Panov, Korneev,
  Martinez-Pinedo, and Thielemann}}]{Panov-2013}
\bibinfo{author}{\bibfnamefont{I.~V.} \bibnamefont{Panov}},
  \bibinfo{author}{\bibfnamefont{I.~Y.} \bibnamefont{Korneev}},
  \bibinfo{author}{\bibfnamefont{G.}~\bibnamefont{Martinez-Pinedo}},
  \bibnamefont{and} \bibinfo{author}{\bibfnamefont{F.-K.}
  \bibnamefont{Thielemann}}, \bibinfo{journal}{Astron. Lett.}
  \textbf{\bibinfo{volume}{39}}, \bibinfo{pages}{150} (\bibinfo{year}{2013}).

\bibitem[{\citenamefont{Erler et~al.}(2012)\citenamefont{Erler, Langanke,
  Loens, Martinez-Pinedo, and Reinhard}}]{Erler-2012}
\bibinfo{author}{\bibfnamefont{J.}~\bibnamefont{Erler}},
  \bibinfo{author}{\bibfnamefont{K.}~\bibnamefont{Langanke}},
  \bibinfo{author}{\bibfnamefont{H.~P.} \bibnamefont{Loens}},
  \bibinfo{author}{\bibfnamefont{G.}~\bibnamefont{Martinez-Pinedo}},
  \bibnamefont{and} \bibinfo{author}{\bibfnamefont{P.~G.}
  \bibnamefont{Reinhard}}, \bibinfo{journal}{Phys. Rev. C}
  \textbf{\bibinfo{volume}{85}}, \bibinfo{pages}{025802}
  (\bibinfo{year}{2012}).

\bibitem[{\citenamefont{Thoennessen and Sherrill}(2011)}]{Thoenn-2011}
\bibinfo{author}{\bibfnamefont{M.}~\bibnamefont{Thoennessen}} \bibnamefont{and}
  \bibinfo{author}{\bibfnamefont{B.}~\bibnamefont{Sherrill}},
  \bibinfo{journal}{Nature} \textbf{\bibinfo{volume}{473}}, \bibinfo{pages}{25}
  (\bibinfo{year}{2011}).

\bibitem[{\citenamefont{Oganessian and Utyonkov}(2015)}]{Oganessian-2015}
\bibinfo{author}{\bibfnamefont{Y.~T.} \bibnamefont{Oganessian}}
  \bibnamefont{and} \bibinfo{author}{\bibfnamefont{V.~K.}
  \bibnamefont{Utyonkov}}, \bibinfo{journal}{Rep. Prog. Phys.}
  \textbf{\bibinfo{volume}{78}}, \bibinfo{pages}{036301}
  (\bibinfo{year}{2015}).

\bibitem[{\citenamefont{Oganessian}(2007)}]{SHE-2014}
\bibinfo{author}{\bibfnamefont{Y.~T.} \bibnamefont{Oganessian}},
  \bibinfo{journal}{J. Phys. G} \textbf{\bibinfo{volume}{34}},
  \bibinfo{pages}{165} (\bibinfo{year}{2007}).

\bibitem[{\citenamefont{Thoennessen}(2013)}]{Thoenn-2013}
\bibinfo{author}{\bibfnamefont{M.}~\bibnamefont{Thoennessen}},
  \bibinfo{journal}{Rep. Prog. Phys.} \textbf{\bibinfo{volume}{76}},
  \bibinfo{pages}{056301} (\bibinfo{year}{2013}).

\bibitem[{\citenamefont{Seaborg and Loveland}(1990)}]{Loveland-1990}
\bibinfo{author}{\bibfnamefont{G.~T.} \bibnamefont{Seaborg}} \bibnamefont{and}
  \bibinfo{author}{\bibfnamefont{W.~D.} \bibnamefont{Loveland}},
  \emph{\bibinfo{title}{The Elements Beyond Uranium}}
  (\bibinfo{publisher}{Wiley}, \bibinfo{year}{1990}).

\bibitem[{\citenamefont{Meitner and Frisch}(1939)}]{Meitner-1939}
\bibinfo{author}{\bibfnamefont{L.}~\bibnamefont{Meitner}} \bibnamefont{and}
  \bibinfo{author}{\bibfnamefont{O.~R.} \bibnamefont{Frisch}},
  \bibinfo{journal}{Nature} \textbf{\bibinfo{volume}{143}},
  \bibinfo{pages}{239} (\bibinfo{year}{1939}).

\bibitem[{\citenamefont{Bohr and Wheeler}(1939)}]{Bohr-1939}
\bibinfo{author}{\bibfnamefont{N.}~\bibnamefont{Bohr}} \bibnamefont{and}
  \bibinfo{author}{\bibfnamefont{J.~A.} \bibnamefont{Wheeler}},
  \bibinfo{journal}{Phys. Rev.} \textbf{\bibinfo{volume}{56}},
  \bibinfo{pages}{426} (\bibinfo{year}{1939}).

\bibitem[{\citenamefont{Frankel and Metropolis}(1947)}]{Frankel-1947}
\bibinfo{author}{\bibfnamefont{S.}~\bibnamefont{Frankel}} \bibnamefont{and}
  \bibinfo{author}{\bibfnamefont{N.}~\bibnamefont{Metropolis}},
  \bibinfo{journal}{Phys. Rev.} \textbf{\bibinfo{volume}{72}},
  \bibinfo{pages}{914} (\bibinfo{year}{1947}).

\bibitem[{\citenamefont{Fong}(1956)}]{Fong-1956}
\bibinfo{author}{\bibfnamefont{P.}~\bibnamefont{Fong}}, \bibinfo{journal}{Phys.
  Rev.} \textbf{\bibinfo{volume}{102}}, \bibinfo{pages}{434}
  (\bibinfo{year}{1956}).

\bibitem[{\citenamefont{B.D.Wilkins et~al.}(1976)\citenamefont{B.D.Wilkins,
  Steinberg, and Chasman}}]{Wilkins-1976}
\bibinfo{author}{\bibnamefont{B.D.Wilkins}},
  \bibinfo{author}{\bibfnamefont{E.}~\bibnamefont{Steinberg}},
  \bibnamefont{and} \bibinfo{author}{\bibfnamefont{R.~R.}
  \bibnamefont{Chasman}}, \bibinfo{journal}{Phys. Rev. C}
  \textbf{\bibinfo{volume}{14}}, \bibinfo{pages}{1832} (\bibinfo{year}{1976}).

\bibitem[{\citenamefont{Brosa et~al.}(2001)\citenamefont{Brosa, Grossman, and
  Muller}}]{Brosa-1990}
\bibinfo{author}{\bibfnamefont{U.}~\bibnamefont{Brosa}},
  \bibinfo{author}{\bibfnamefont{S.}~\bibnamefont{Grossman}}, \bibnamefont{and}
  \bibinfo{author}{\bibfnamefont{A.}~\bibnamefont{Muller}},
  \bibinfo{journal}{Phys. Rep.} \textbf{\bibinfo{volume}{167}},
  \bibinfo{pages}{1990} (\bibinfo{year}{2001}).

\bibitem[{\citenamefont{Duijvestijn et~al.}(2001)\citenamefont{Duijvestijn,
  Koning, and Hambsch}}]{Dui-2001}
\bibinfo{author}{\bibfnamefont{M.~C.} \bibnamefont{Duijvestijn}},
  \bibinfo{author}{\bibfnamefont{A.~J.} \bibnamefont{Koning}},
  \bibnamefont{and} \bibinfo{author}{\bibfnamefont{F.~J.}
  \bibnamefont{Hambsch}}, \bibinfo{journal}{Phys. Rev. C}
  \textbf{\bibinfo{volume}{64}}, \bibinfo{pages}{014607}
  (\bibinfo{year}{2001}).

\bibitem[{\citenamefont{Demetriou et~al.}(2010)\citenamefont{Demetriou,
  Keutgen, Prieels, and ElMasri}}]{Demetriou-2010}
\bibinfo{author}{\bibfnamefont{P.}~\bibnamefont{Demetriou}},
  \bibinfo{author}{\bibfnamefont{T.}~\bibnamefont{Keutgen}},
  \bibinfo{author}{\bibfnamefont{R.}~\bibnamefont{Prieels}}, \bibnamefont{and}
  \bibinfo{author}{\bibfnamefont{Y.}~\bibnamefont{ElMasri}},
  \bibinfo{journal}{Phys. Rev. C} \textbf{\bibinfo{volume}{82}},
  \bibinfo{pages}{054606} (\bibinfo{year}{2010}).

\bibitem[{\citenamefont{Ryzhov et~al.}(2011)\citenamefont{Ryzhov, Yavshits,
  Tutin et~al.}}]{Ryzhov-2011}
\bibinfo{author}{\bibfnamefont{I.~V.} \bibnamefont{Ryzhov}},
  \bibinfo{author}{\bibfnamefont{S.~G.} \bibnamefont{Yavshits}},
  \bibinfo{author}{\bibfnamefont{G.~A.} \bibnamefont{Tutin}},
  \bibnamefont{et~al.}, \bibinfo{journal}{Phys. Rev. C}
  \textbf{\bibinfo{volume}{83}}, \bibinfo{pages}{054603}
  (\bibinfo{year}{2011}).

\bibitem[{\citenamefont{Rubchenya and Aysto}(2012)}]{Rubchenya-2012}
\bibinfo{author}{\bibfnamefont{V.~A.} \bibnamefont{Rubchenya}}
  \bibnamefont{and} \bibinfo{author}{\bibfnamefont{J.}~\bibnamefont{Aysto}},
  \bibinfo{journal}{Eur. Phys. J. A} \textbf{\bibinfo{volume}{48}},
  \bibinfo{pages}{44} (\bibinfo{year}{2012}).

\bibitem[{\citenamefont{Maslov}(2003)}]{Maslov-2003}
\bibinfo{author}{\bibfnamefont{V.~M.} \bibnamefont{Maslov}},
  \bibinfo{journal}{Nucl. Phys. A} \textbf{\bibinfo{volume}{717}},
  \bibinfo{pages}{3} (\bibinfo{year}{2003}).

\bibitem[{\citenamefont{Lestone and McCalla}(2009)}]{Lestone-2009}
\bibinfo{author}{\bibfnamefont{J.~P.} \bibnamefont{Lestone}} \bibnamefont{and}
  \bibinfo{author}{\bibfnamefont{S.~G.} \bibnamefont{McCalla}},
  \bibinfo{journal}{Phys. Rev. C} \textbf{\bibinfo{volume}{79}},
  \bibinfo{pages}{044611} (\bibinfo{year}{2009}).

\bibitem[{\citenamefont{Charity}(2010)}]{GEMINI-2010}
\bibinfo{author}{\bibfnamefont{R.~J.} \bibnamefont{Charity}},
  \bibinfo{journal}{Phys. Rev. C} \textbf{\bibinfo{volume}{82}},
  \bibinfo{pages}{014610} (\bibinfo{year}{2010}).

\bibitem[{\citenamefont{Moretto}(1979)}]{Moretto}
\bibinfo{author}{\bibfnamefont{L.}~\bibnamefont{Moretto}},
  \bibinfo{journal}{Nucl. Phys. A} \textbf{\bibinfo{volume}{00}},
  \bibinfo{pages}{0} (\bibinfo{year}{1979}).

\bibitem[{\citenamefont{Eren et~al.}(2013)\citenamefont{Eren, Buyukcizmeci,
  Ogul, and Botvina}}]{SMM-2013}
\bibinfo{author}{\bibfnamefont{N.}~\bibnamefont{Eren}},
  \bibinfo{author}{\bibfnamefont{N.}~\bibnamefont{Buyukcizmeci}},
  \bibinfo{author}{\bibfnamefont{R.}~\bibnamefont{Ogul}}, \bibnamefont{and}
  \bibinfo{author}{\bibfnamefont{A.~S.} \bibnamefont{Botvina}},
  \bibinfo{journal}{Eur. Phys. J. A} \textbf{\bibinfo{volume}{49}},
  \bibinfo{pages}{48} (\bibinfo{year}{2013}).

\bibitem[{\citenamefont{Kramers}(1940)}]{Kramers-1940}
\bibinfo{author}{\bibfnamefont{H.~A.} \bibnamefont{Kramers}},
  \bibinfo{journal}{Physica (Amsterdam)} \textbf{\bibinfo{volume}{7}},
  \bibinfo{pages}{284} (\bibinfo{year}{1940}).

\bibitem[{\citenamefont{Abe et~al.}(1996)\citenamefont{Abe, Ayik, Reinhard, and
  Suraud}}]{Abe-1996}
\bibinfo{author}{\bibfnamefont{Y.}~\bibnamefont{Abe}},
  \bibinfo{author}{\bibfnamefont{S.}~\bibnamefont{Ayik}},
  \bibinfo{author}{\bibfnamefont{P.~G.} \bibnamefont{Reinhard}},
  \bibnamefont{and} \bibinfo{author}{\bibfnamefont{E.}~\bibnamefont{Suraud}},
  \bibinfo{journal}{Phys. Rep.} \textbf{\bibinfo{volume}{275}},
  \bibinfo{pages}{49} (\bibinfo{year}{1996}).

\bibitem[{\citenamefont{Frobrich and Gontchar}(1998)}]{Frobrich-1998}
\bibinfo{author}{\bibfnamefont{P.}~\bibnamefont{Frobrich}} \bibnamefont{and}
  \bibinfo{author}{\bibfnamefont{I.~I.} \bibnamefont{Gontchar}},
  \bibinfo{journal}{Phys. Rep.} \textbf{\bibinfo{volume}{292}},
  \bibinfo{pages}{131} (\bibinfo{year}{1998}).

\bibitem[{\citenamefont{Chaudhuri and Pal}(2001)}]{Chaudhuri-2001}
\bibinfo{author}{\bibfnamefont{G.}~\bibnamefont{Chaudhuri}} \bibnamefont{and}
  \bibinfo{author}{\bibfnamefont{S.}~\bibnamefont{Pal}},
  \bibinfo{journal}{Phys. Rev. C} \textbf{\bibinfo{volume}{63}},
  \bibinfo{pages}{064603} (\bibinfo{year}{2001}).

\bibitem[{\citenamefont{Karpov et~al.}(2001)\citenamefont{Karpov, Nadtochy,
  Vanin, and Adeev}}]{Karpov-2001}
\bibinfo{author}{\bibfnamefont{A.~V.} \bibnamefont{Karpov}},
  \bibinfo{author}{\bibfnamefont{P.~N.} \bibnamefont{Nadtochy}},
  \bibinfo{author}{\bibfnamefont{D.~V.} \bibnamefont{Vanin}}, \bibnamefont{and}
  \bibinfo{author}{\bibfnamefont{G.~D.} \bibnamefont{Adeev}},
  \bibinfo{journal}{Phys. Rev. C} \textbf{\bibinfo{volume}{63}},
  \bibinfo{pages}{054610} (\bibinfo{year}{2001}).

\bibitem[{\citenamefont{Nadtochy et~al.}(2007)\citenamefont{Nadtochy,
  Keli\'{c}, and Schmidt}}]{Nadtochy-2007}
\bibinfo{author}{\bibfnamefont{P.~N.} \bibnamefont{Nadtochy}},
  \bibinfo{author}{\bibfnamefont{A.}~\bibnamefont{Keli\'{c}}},
  \bibnamefont{and} \bibinfo{author}{\bibfnamefont{K.~H.}
  \bibnamefont{Schmidt}}, \bibinfo{journal}{Phys. Rev. C}
  \textbf{\bibinfo{volume}{75}}, \bibinfo{pages}{064614}
  (\bibinfo{year}{2007}).

\bibitem[{\citenamefont{Sadhukhan and Pal}(2009)}]{Sad-2009}
\bibinfo{author}{\bibfnamefont{J.}~\bibnamefont{Sadhukhan}} \bibnamefont{and}
  \bibinfo{author}{\bibfnamefont{S.}~\bibnamefont{Pal}},
  \bibinfo{journal}{Phys. Rev. C} \textbf{\bibinfo{volume}{79}},
  \bibinfo{pages}{064606} (\bibinfo{year}{2009}).

\bibitem[{\citenamefont{Sadhukhan and Pal}(2011)}]{Sad-2011}
\bibinfo{author}{\bibfnamefont{J.}~\bibnamefont{Sadhukhan}} \bibnamefont{and}
  \bibinfo{author}{\bibfnamefont{S.}~\bibnamefont{Pal}},
  \bibinfo{journal}{Phys. Rev. C} \textbf{\bibinfo{volume}{84}},
  \bibinfo{pages}{044610} (\bibinfo{year}{2011}).

\bibitem[{\citenamefont{Nadtochy et~al.}(2012)\citenamefont{Nadtochy, Ryabov,
  Gegechkori, Anischenko, and Adeev}}]{Nadtochy-2012}
\bibinfo{author}{\bibfnamefont{P.~N.} \bibnamefont{Nadtochy}},
  \bibinfo{author}{\bibfnamefont{E.~G.} \bibnamefont{Ryabov}},
  \bibinfo{author}{\bibfnamefont{A.~E.} \bibnamefont{Gegechkori}},
  \bibinfo{author}{\bibfnamefont{Y.~A.} \bibnamefont{Anischenko}},
  \bibnamefont{and} \bibinfo{author}{\bibfnamefont{G.~D.} \bibnamefont{Adeev}},
  \bibinfo{journal}{Phys. Rev. C} \textbf{\bibinfo{volume}{85}},
  \bibinfo{pages}{064619} (\bibinfo{year}{2012}).

\bibitem[{\citenamefont{Eslamizadeh}(2014)}]{Eslam-2014}
\bibinfo{author}{\bibfnamefont{H.}~\bibnamefont{Eslamizadeh}},
  \bibinfo{journal}{Eur. Phys. J. A} \textbf{\bibinfo{volume}{50}}
  (\bibinfo{year}{2014}).

\bibitem[{\citenamefont{Ye et~al.}(2014)\citenamefont{Ye, Wang, and
  Tian}}]{Ye-2014}
\bibinfo{author}{\bibfnamefont{W.}~\bibnamefont{Ye}},
  \bibinfo{author}{\bibfnamefont{N.}~\bibnamefont{Wang}}, \bibnamefont{and}
  \bibinfo{author}{\bibfnamefont{J.}~\bibnamefont{Tian}},
  \bibinfo{journal}{Phys. Rev. C} \textbf{\bibinfo{volume}{90}},
  \bibinfo{pages}{041604} (\bibinfo{year}{2014}).

\bibitem[{\citenamefont{Aritomo et~al.}(2014)\citenamefont{Aritomo, Chiba, and
  Ivanyuk}}]{Aritomo-2014a}
\bibinfo{author}{\bibfnamefont{Y.}~\bibnamefont{Aritomo}},
  \bibinfo{author}{\bibfnamefont{S.}~\bibnamefont{Chiba}}, \bibnamefont{and}
  \bibinfo{author}{\bibfnamefont{F.~A.} \bibnamefont{Ivanyuk}},
  \bibinfo{journal}{Phys. Rev. C} \textbf{\bibinfo{volume}{90}},
  \bibinfo{pages}{054609} (\bibinfo{year}{2014}).

\bibitem[{\citenamefont{Ivanyuk et~al.}(2014)\citenamefont{Ivanyuk, Chiba, and
  Aritomo}}]{Aritomo-2014b}
\bibinfo{author}{\bibfnamefont{F.~A.} \bibnamefont{Ivanyuk}},
  \bibinfo{author}{\bibfnamefont{S.}~\bibnamefont{Chiba}}, \bibnamefont{and}
  \bibinfo{author}{\bibfnamefont{Y.}~\bibnamefont{Aritomo}},
  \bibinfo{journal}{Phys. Rev. C} \textbf{\bibinfo{volume}{90}},
  \bibinfo{pages}{054607} (\bibinfo{year}{2014}).

\bibitem[{\citenamefont{Nadtochy et~al.}(2014)\citenamefont{Nadtochy, Ryabov,
  Gegechkori, Anischenko, and Adeev}}]{Nadtochy-2014}
\bibinfo{author}{\bibfnamefont{P.~N.} \bibnamefont{Nadtochy}},
  \bibinfo{author}{\bibfnamefont{E.~G.} \bibnamefont{Ryabov}},
  \bibinfo{author}{\bibfnamefont{A.~E.} \bibnamefont{Gegechkori}},
  \bibinfo{author}{\bibfnamefont{Y.~A.} \bibnamefont{Anischenko}},
  \bibnamefont{and} \bibinfo{author}{\bibfnamefont{G.~D.} \bibnamefont{Adeev}},
  \bibinfo{journal}{Phys. Rev. C} \textbf{\bibinfo{volume}{89}},
  \bibinfo{pages}{014616} (\bibinfo{year}{2014}).

\bibitem[{\citenamefont{Aritomo and Chiba}(2013)}]{Aritomo-2013}
\bibinfo{author}{\bibfnamefont{Y.~A.} \bibnamefont{Aritomo}} \bibnamefont{and}
  \bibinfo{author}{\bibfnamefont{S.}~\bibnamefont{Chiba}},
  \bibinfo{journal}{Phys. Rev. C} \textbf{\bibinfo{volume}{88}},
  \bibinfo{pages}{044614} (\bibinfo{year}{2013}).

\bibitem[{\citenamefont{Randrup and M\"{o}ller}(2013)}]{Randrup-2013}
\bibinfo{author}{\bibfnamefont{J.}~\bibnamefont{Randrup}} \bibnamefont{and}
  \bibinfo{author}{\bibfnamefont{P.}~\bibnamefont{M\"{o}ller}},
  \bibinfo{journal}{Phys. Rev. C} \textbf{\bibinfo{volume}{88}},
  \bibinfo{pages}{064606} (\bibinfo{year}{2013}).

\bibitem[{\citenamefont{Randrup et~al.}(2011)\citenamefont{Randrup, M\"{o}ller,
  and Sierk}}]{Randrup-2011a}
\bibinfo{author}{\bibfnamefont{J.}~\bibnamefont{Randrup}},
  \bibinfo{author}{\bibfnamefont{P.}~\bibnamefont{M\"{o}ller}},
  \bibnamefont{and} \bibinfo{author}{\bibfnamefont{A.~J.} \bibnamefont{Sierk}},
  \bibinfo{journal}{Phys. Rev. C} \textbf{\bibinfo{volume}{84}},
  \bibinfo{pages}{034613} (\bibinfo{year}{2011}).

\bibitem[{\citenamefont{Randrup and M\"{o}ller}(2011)}]{Randrup-2011b}
\bibinfo{author}{\bibfnamefont{J.}~\bibnamefont{Randrup}} \bibnamefont{and}
  \bibinfo{author}{\bibfnamefont{P.}~\bibnamefont{M\"{o}ller}},
  \bibinfo{journal}{Phys. Rev. Lett.} \textbf{\bibinfo{volume}{106}},
  \bibinfo{pages}{132503} (\bibinfo{year}{2011}).

\bibitem[{\citenamefont{M\"{o}ller and Randrup}(2015)}]{Moller-2015z}
\bibinfo{author}{\bibfnamefont{P.}~\bibnamefont{M\"{o}ller}} \bibnamefont{and}
  \bibinfo{author}{\bibfnamefont{J.}~\bibnamefont{Randrup}},
  \bibinfo{journal}{Phys. Rev. C} \textbf{\bibinfo{volume}{91}},
  \bibinfo{pages}{044316} (\bibinfo{year}{2015}).

\bibitem[{\citenamefont{Mazurek et~al.}(2015)\citenamefont{Mazurek, Schmitt,
  and Nadtochy}}]{Mazurek-2015z}
\bibinfo{author}{\bibfnamefont{K.}~\bibnamefont{Mazurek}},
  \bibinfo{author}{\bibfnamefont{C.}~\bibnamefont{Schmitt}}, \bibnamefont{and}
  \bibinfo{author}{\bibfnamefont{P.~N.} \bibnamefont{Nadtochy}},
  \bibinfo{journal}{Phys. Rev. C} \textbf{\bibinfo{volume}{91}},
  \bibinfo{pages}{041603} (\bibinfo{year}{2015}).

\bibitem[{\citenamefont{Swiatecki}(1955)}]{Swiatecki-1955}
\bibinfo{author}{\bibfnamefont{W.~J.} \bibnamefont{Swiatecki}},
  \bibinfo{journal}{Phys. Rev.} \textbf{\bibinfo{volume}{100}},
  \bibinfo{pages}{937} (\bibinfo{year}{1955}).

\bibitem[{\citenamefont{Strutinsky}(1967)}]{Strutinsky-1967}
\bibinfo{author}{\bibfnamefont{V.~M.} \bibnamefont{Strutinsky}},
  \bibinfo{journal}{Nucl. Phys. A} \textbf{\bibinfo{volume}{95}},
  \bibinfo{pages}{420} (\bibinfo{year}{1967}).

\bibitem[{\citenamefont{Strutinsky}(1968)}]{Strutinsky-1968}
\bibinfo{author}{\bibfnamefont{V.~M.} \bibnamefont{Strutinsky}},
  \bibinfo{journal}{Nucl. Phys. A} \textbf{\bibinfo{volume}{122}},
  \bibinfo{pages}{1} (\bibinfo{year}{1968}).

\bibitem[{\citenamefont{M\"{o}ller et~al.}(2009)\citenamefont{M\"{o}ller,
  Sierk, Ichikawa et~al.}}]{Moller-2009}
\bibinfo{author}{\bibfnamefont{P.}~\bibnamefont{M\"{o}ller}},
  \bibinfo{author}{\bibfnamefont{A.~J.} \bibnamefont{Sierk}},
  \bibinfo{author}{\bibfnamefont{T.}~\bibnamefont{Ichikawa}},
  \bibnamefont{et~al.}, \bibinfo{journal}{Phys. Rev. C}
  \textbf{\bibinfo{volume}{79}}, \bibinfo{pages}{064304}
  (\bibinfo{year}{2009}).

\bibitem[{\citenamefont{Pomorski}(2013)}]{Pomorski-2013}
\bibinfo{author}{\bibfnamefont{K.}~\bibnamefont{Pomorski}},
  \bibinfo{journal}{Phys. Scr. T} \textbf{\bibinfo{volume}{154}},
  \bibinfo{pages}{014023} (\bibinfo{year}{2013}).

\bibitem[{\citenamefont{M\"{o}ller et~al.}(2015)\citenamefont{M\"{o}ller,
  Sierk, Ichikawa, Iwamoto, and Mumpower}}]{Moller-2015}
\bibinfo{author}{\bibfnamefont{P.}~\bibnamefont{M\"{o}ller}},
  \bibinfo{author}{\bibfnamefont{A.~J.} \bibnamefont{Sierk}},
  \bibinfo{author}{\bibfnamefont{T.}~\bibnamefont{Ichikawa}},
  \bibinfo{author}{\bibfnamefont{A.}~\bibnamefont{Iwamoto}}, \bibnamefont{and}
  \bibinfo{author}{\bibfnamefont{M.}~\bibnamefont{Mumpower}},
  \bibinfo{journal}{Phys. Rev. C} \textbf{\bibinfo{volume}{91}},
  \bibinfo{pages}{024310} (\bibinfo{year}{2015}).

\bibitem[{\citenamefont{M\"{o}ller et~al.}(2001)\citenamefont{M\"{o}ller,
  Madland, Sierk, and Iwamoto}}]{Moller-2001}
\bibinfo{author}{\bibfnamefont{P.}~\bibnamefont{M\"{o}ller}},
  \bibinfo{author}{\bibfnamefont{D.~G.} \bibnamefont{Madland}},
  \bibinfo{author}{\bibfnamefont{A.~J.} \bibnamefont{Sierk}}, \bibnamefont{and}
  \bibinfo{author}{\bibfnamefont{A.}~\bibnamefont{Iwamoto}},
  \bibinfo{journal}{Nature} \textbf{\bibinfo{volume}{409}},
  \bibinfo{pages}{785} (\bibinfo{year}{2001}).

\bibitem[{\citenamefont{Rodriguez-Guzm\'{a}n and
  Robledo}(2014)}]{Rodriguez-2014}
\bibinfo{author}{\bibfnamefont{R.}~\bibnamefont{Rodriguez-Guzm\'{a}n}}
  \bibnamefont{and} \bibinfo{author}{\bibfnamefont{L.~M.}
  \bibnamefont{Robledo}}, \bibinfo{journal}{Phys. Rev. C}
  \textbf{\bibinfo{volume}{89}}, \bibinfo{pages}{054310}
  (\bibinfo{year}{2014}).

\bibitem[{\citenamefont{Sadhukhan et~al.}(2014)\citenamefont{Sadhukhan,
  Dobaczewski, Nazarewicz, Sheikh, and Baran}}]{Sad-2014}
\bibinfo{author}{\bibfnamefont{J.}~\bibnamefont{Sadhukhan}},
  \bibinfo{author}{\bibfnamefont{J.}~\bibnamefont{Dobaczewski}},
  \bibinfo{author}{\bibfnamefont{W.}~\bibnamefont{Nazarewicz}},
  \bibinfo{author}{\bibfnamefont{J.~A.} \bibnamefont{Sheikh}},
  \bibnamefont{and} \bibinfo{author}{\bibfnamefont{A.}~\bibnamefont{Baran}},
  \bibinfo{journal}{Phys. Rev. C} \textbf{\bibinfo{volume}{90}},
  \bibinfo{pages}{061304} (\bibinfo{year}{2014}).

\bibitem[{\citenamefont{Sadhukhan et~al.}(2013)\citenamefont{Sadhukhan,
  Mazurek, Baran, Dobaczewski, Nazarewicz, and Sheikh}}]{Sad-2013}
\bibinfo{author}{\bibfnamefont{J.}~\bibnamefont{Sadhukhan}},
  \bibinfo{author}{\bibfnamefont{K.}~\bibnamefont{Mazurek}},
  \bibinfo{author}{\bibfnamefont{A.}~\bibnamefont{Baran}},
  \bibinfo{author}{\bibfnamefont{J.}~\bibnamefont{Dobaczewski}},
  \bibinfo{author}{\bibfnamefont{W.}~\bibnamefont{Nazarewicz}},
  \bibnamefont{and} \bibinfo{author}{\bibfnamefont{J.~A.}
  \bibnamefont{Sheikh}}, \bibinfo{journal}{Phys. Rev. C}
  \textbf{\bibinfo{volume}{88}}, \bibinfo{pages}{064314}
  (\bibinfo{year}{2013}).

\bibitem[{\citenamefont{Giuliani
  et~al.}(2014{\natexlab{a}})\citenamefont{Giuliani, Robledo, and
  Rodriguez-Guzman}}]{Giuliani-2014}
\bibinfo{author}{\bibfnamefont{S.~A.} \bibnamefont{Giuliani}},
  \bibinfo{author}{\bibfnamefont{L.~M.} \bibnamefont{Robledo}},
  \bibnamefont{and}
  \bibinfo{author}{\bibfnamefont{R.}~\bibnamefont{Rodriguez-Guzman}},
  \bibinfo{journal}{Phys. Rev. C} \textbf{\bibinfo{volume}{90}},
  \bibinfo{pages}{054311} (\bibinfo{year}{2014}{\natexlab{a}}).

\bibitem[{\citenamefont{Schunck et~al.}(2014)\citenamefont{Schunck, Duke, Carr,
  and Knoll}}]{Schunck-2014}
\bibinfo{author}{\bibfnamefont{N.}~\bibnamefont{Schunck}},
  \bibinfo{author}{\bibfnamefont{D.}~\bibnamefont{Duke}},
  \bibinfo{author}{\bibfnamefont{H.}~\bibnamefont{Carr}}, \bibnamefont{and}
  \bibinfo{author}{\bibfnamefont{A.}~\bibnamefont{Knoll}},
  \bibinfo{journal}{Phys. Rev. C} \textbf{\bibinfo{volume}{90}},
  \bibinfo{pages}{054305} (\bibinfo{year}{2014}).

\bibitem[{\citenamefont{Schunck et~al.}(2015)\citenamefont{Schunck, Duke, and
  Carr}}]{Schunck-2015}
\bibinfo{author}{\bibfnamefont{N.}~\bibnamefont{Schunck}},
  \bibinfo{author}{\bibfnamefont{D.}~\bibnamefont{Duke}}, \bibnamefont{and}
  \bibinfo{author}{\bibfnamefont{H.}~\bibnamefont{Carr}},
  \bibinfo{journal}{Phys. Rev. C} \textbf{\bibinfo{volume}{91}},
  \bibinfo{pages}{034327} (\bibinfo{year}{2015}).

\bibitem[{\citenamefont{Giuliani and Robledo}(2013)}]{Giuliani-2013}
\bibinfo{author}{\bibfnamefont{S.~A.} \bibnamefont{Giuliani}} \bibnamefont{and}
  \bibinfo{author}{\bibfnamefont{L.~M.} \bibnamefont{Robledo}},
  \bibinfo{journal}{Phys. Rev. C} \textbf{\bibinfo{volume}{88}},
  \bibinfo{pages}{054325} (\bibinfo{year}{2013}).

\bibitem[{McD()}]{McDonnell-2012}
\bibinfo{note}{J. D. McDonnell, PhD Dissertation, University of Tennesse
  (2012), http://trace.tennessee.edu/utk$_{-}$graddiss/1438}.

\bibitem[{\citenamefont{McDonnell et~al.}(2014)\citenamefont{McDonnell,
  Nazarewicz, Sheikh, Staszczak, and Warda}}]{McDonnell-2014}
\bibinfo{author}{\bibfnamefont{J.~D.} \bibnamefont{McDonnell}},
  \bibinfo{author}{\bibfnamefont{W.}~\bibnamefont{Nazarewicz}},
  \bibinfo{author}{\bibfnamefont{J.~A.} \bibnamefont{Sheikh}},
  \bibinfo{author}{\bibfnamefont{A.}~\bibnamefont{Staszczak}},
  \bibnamefont{and} \bibinfo{author}{\bibfnamefont{M.}~\bibnamefont{Warda}},
  \bibinfo{journal}{Phys. Rev. C} \textbf{\bibinfo{volume}{90}},
  \bibinfo{pages}{021302} (\bibinfo{year}{2014}).

\bibitem[{\citenamefont{Pei et~al.}(2014)\citenamefont{Pei, Fann, Harrison,
  Nazarewicz, Shi, and Thornton}}]{Pei-2014}
\bibinfo{author}{\bibfnamefont{J.~C.} \bibnamefont{Pei}},
  \bibinfo{author}{\bibfnamefont{G.~I.} \bibnamefont{Fann}},
  \bibinfo{author}{\bibfnamefont{R.~J.} \bibnamefont{Harrison}},
  \bibinfo{author}{\bibfnamefont{W.}~\bibnamefont{Nazarewicz}},
  \bibinfo{author}{\bibfnamefont{Y.}~\bibnamefont{Shi}}, \bibnamefont{and}
  \bibinfo{author}{\bibfnamefont{S.}~\bibnamefont{Thornton}},
  \bibinfo{journal}{Phys. Rev. C} \textbf{\bibinfo{volume}{90}},
  \bibinfo{pages}{024317} (\bibinfo{year}{2014}).

\bibitem[{\citenamefont{Negele et~al.}(1978)\citenamefont{Negele, Koonin,
  M\'{o}ller, Nix, and Sierk}}]{Negele-1978}
\bibinfo{author}{\bibfnamefont{J.~W.} \bibnamefont{Negele}},
  \bibinfo{author}{\bibfnamefont{S.~E.} \bibnamefont{Koonin}},
  \bibinfo{author}{\bibfnamefont{P.}~\bibnamefont{M\'{o}ller}},
  \bibinfo{author}{\bibfnamefont{J.~R.} \bibnamefont{Nix}}, \bibnamefont{and}
  \bibinfo{author}{\bibfnamefont{A.~J.} \bibnamefont{Sierk}},
  \bibinfo{journal}{Phys. Rev. C} \textbf{\bibinfo{volume}{17}},
  \bibinfo{pages}{1098} (\bibinfo{year}{1978}).

\bibitem[{\citenamefont{Sekizawa and Yabana}(2013)}]{TDHF-HI-2013}
\bibinfo{author}{\bibfnamefont{K.}~\bibnamefont{Sekizawa}} \bibnamefont{and}
  \bibinfo{author}{\bibfnamefont{K.}~\bibnamefont{Yabana}},
  \bibinfo{journal}{Phys. Rev. C} \textbf{\bibinfo{volume}{88}},
  \bibinfo{pages}{014614} (\bibinfo{year}{2013}).

\bibitem[{\citenamefont{Yilmaz et~al.}(2011)\citenamefont{Yilmaz, Ayik,
  Lacroix, and Washiyama}}]{TDHF-HI-2011}
\bibinfo{author}{\bibfnamefont{B.}~\bibnamefont{Yilmaz}},
  \bibinfo{author}{\bibfnamefont{S.}~\bibnamefont{Ayik}},
  \bibinfo{author}{\bibfnamefont{D.}~\bibnamefont{Lacroix}}, \bibnamefont{and}
  \bibinfo{author}{\bibfnamefont{K.}~\bibnamefont{Washiyama}},
  \bibinfo{journal}{Phys. Rev. C} \textbf{\bibinfo{volume}{83}},
  \bibinfo{pages}{064615} (\bibinfo{year}{2011}).

\bibitem[{\citenamefont{Iwata et~al.}(2010)\citenamefont{Iwata, Otsuka, Maruhn,
  and Itagaki}}]{TDHF-HI-2010}
\bibinfo{author}{\bibfnamefont{Y.}~\bibnamefont{Iwata}},
  \bibinfo{author}{\bibfnamefont{T.}~\bibnamefont{Otsuka}},
  \bibinfo{author}{\bibfnamefont{J.~A.} \bibnamefont{Maruhn}},
  \bibnamefont{and} \bibinfo{author}{\bibfnamefont{N.}~\bibnamefont{Itagaki}},
  \bibinfo{journal}{Phys. Rev. Lett.} \textbf{\bibinfo{volume}{104}},
  \bibinfo{pages}{252501} (\bibinfo{year}{2010}).

\bibitem[{\citenamefont{Baranger and V\'{e}n\'{e}roni}(1978)}]{Baranger-1978}
\bibinfo{author}{\bibfnamefont{M.}~\bibnamefont{Baranger}} \bibnamefont{and}
  \bibinfo{author}{\bibfnamefont{M.}~\bibnamefont{V\'{e}n\'{e}roni}},
  \bibinfo{journal}{Ann. Phys. (NY)} \textbf{\bibinfo{volume}{114}},
  \bibinfo{pages}{123} (\bibinfo{year}{1978}).

\bibitem[{\citenamefont{Goutte et~al.}(2005)\citenamefont{Goutte, Berger,
  Casoli, and Gogny}}]{Goutte-2005}
\bibinfo{author}{\bibfnamefont{H.}~\bibnamefont{Goutte}},
  \bibinfo{author}{\bibfnamefont{J.~F.} \bibnamefont{Berger}},
  \bibinfo{author}{\bibfnamefont{P.}~\bibnamefont{Casoli}}, \bibnamefont{and}
  \bibinfo{author}{\bibfnamefont{D.}~\bibnamefont{Gogny}},
  \bibinfo{journal}{Phys. Rev. C} \textbf{\bibinfo{volume}{71}},
  \bibinfo{pages}{024316} (\bibinfo{year}{2005}).

\bibitem[{\citenamefont{Rizea and Carjan}(2013)}]{Rizea-2013}
\bibinfo{author}{\bibfnamefont{M.}~\bibnamefont{Rizea}} \bibnamefont{and}
  \bibinfo{author}{\bibfnamefont{N.}~\bibnamefont{Carjan}},
  \bibinfo{journal}{Nucl. Phys. A} \textbf{\bibinfo{volume}{909}},
  \bibinfo{pages}{50} (\bibinfo{year}{2013}).

\bibitem[{\citenamefont{Simenel and Umar}(2014)}]{Simenel-2014}
\bibinfo{author}{\bibfnamefont{C.}~\bibnamefont{Simenel}} \bibnamefont{and}
  \bibinfo{author}{\bibfnamefont{A.~S.} \bibnamefont{Umar}},
  \bibinfo{journal}{Phys. Rev. C} \textbf{\bibinfo{volume}{89}},
  \bibinfo{pages}{031601} (\bibinfo{year}{2014}).

\bibitem[{\citenamefont{Negele}(1989)}]{Negele-1989}
\bibinfo{author}{\bibfnamefont{J.~W.} \bibnamefont{Negele}},
  \bibinfo{journal}{Nucl. Phys. A} \textbf{\bibinfo{volume}{502}},
  \bibinfo{pages}{371} (\bibinfo{year}{1989}).

\bibitem[{\citenamefont{Bonasera and Iwamoto}(1997)}]{Bonasera-1997a}
\bibinfo{author}{\bibfnamefont{A.}~\bibnamefont{Bonasera}} \bibnamefont{and}
  \bibinfo{author}{\bibfnamefont{A.}~\bibnamefont{Iwamoto}},
  \bibinfo{journal}{Phys. Rev. Lett.} \textbf{\bibinfo{volume}{78}},
  \bibinfo{pages}{187} (\bibinfo{year}{1997}).

\bibitem[{\citenamefont{Bonasera et~al.}(1997)\citenamefont{Bonasera,
  Kondratyev, and Iwamoto}}]{Bonasera-1997b}
\bibinfo{author}{\bibfnamefont{A.}~\bibnamefont{Bonasera}},
  \bibinfo{author}{\bibfnamefont{V.~N.} \bibnamefont{Kondratyev}},
  \bibnamefont{and} \bibinfo{author}{\bibfnamefont{A.}~\bibnamefont{Iwamoto}},
  \bibinfo{journal}{J. Phys. G} \textbf{\bibinfo{volume}{23}},
  \bibinfo{pages}{1297} (\bibinfo{year}{1997}).

\bibitem[{\citenamefont{Papa et~al.}(2001)\citenamefont{Papa, Maruyama, and
  Bonasera}}]{Papa-2001}
\bibinfo{author}{\bibfnamefont{M.}~\bibnamefont{Papa}},
  \bibinfo{author}{\bibfnamefont{T.}~\bibnamefont{Maruyama}}, \bibnamefont{and}
  \bibinfo{author}{\bibfnamefont{A.}~\bibnamefont{Bonasera}},
  \bibinfo{journal}{Phys. Rev. C} \textbf{\bibinfo{volume}{64}},
  \bibinfo{pages}{024612} (\bibinfo{year}{2001}).

\bibitem[{\citenamefont{Maruyama et~al.}(2002)\citenamefont{Maruyama, Bonasera,
  Papa, and Chiba}}]{Maru-2002}
\bibinfo{author}{\bibfnamefont{T.}~\bibnamefont{Maruyama}},
  \bibinfo{author}{\bibfnamefont{A.}~\bibnamefont{Bonasera}},
  \bibinfo{author}{\bibfnamefont{M.}~\bibnamefont{Papa}}, \bibnamefont{and}
  \bibinfo{author}{\bibfnamefont{S.}~\bibnamefont{Chiba}},
  \bibinfo{journal}{Eur. Phys. J. A} \textbf{\bibinfo{volume}{14}},
  \bibinfo{pages}{191} (\bibinfo{year}{2002}).

\bibitem[{\citenamefont{Papa et~al.}(2005)\citenamefont{Papa, Giluilani, and
  Bonasera}}]{Papa-2005}
\bibinfo{author}{\bibfnamefont{M.}~\bibnamefont{Papa}},
  \bibinfo{author}{\bibfnamefont{G.}~\bibnamefont{Giluilani}},
  \bibnamefont{and} \bibinfo{author}{\bibfnamefont{A.}~\bibnamefont{Bonasera}},
  \bibinfo{journal}{J Comp. Phys.} \textbf{\bibinfo{volume}{208}},
  \bibinfo{pages}{403} (\bibinfo{year}{2005}).

\bibitem[{\citenamefont{Aichelin}(1991)}]{QMD-1991}
\bibinfo{author}{\bibfnamefont{J.}~\bibnamefont{Aichelin}},
  \bibinfo{journal}{Phys. Rep.} \textbf{\bibinfo{volume}{202}},
  \bibinfo{pages}{233} (\bibinfo{year}{1991}).

\bibitem[{\citenamefont{Papa}(2013)}]{Papa-2013}
\bibinfo{author}{\bibfnamefont{M.}~\bibnamefont{Papa}}, \bibinfo{journal}{Phys.
  Rev. C} \textbf{\bibinfo{volume}{87}}, \bibinfo{pages}{014001}
  (\bibinfo{year}{2013}).

\bibitem[{\citenamefont{Bonasera et~al.}(1994)\citenamefont{Bonasera,
  Gulminelli, and Molitoris}}]{Bonasera-1994}
\bibinfo{author}{\bibfnamefont{A.}~\bibnamefont{Bonasera}},
  \bibinfo{author}{\bibfnamefont{F.}~\bibnamefont{Gulminelli}},
  \bibnamefont{and}
  \bibinfo{author}{\bibfnamefont{J.}~\bibnamefont{Molitoris}},
  \bibinfo{journal}{Phys. Rep.} \textbf{\bibinfo{volume}{243}},
  \bibinfo{pages}{1} (\bibinfo{year}{1994}).

\bibitem[{\citenamefont{Isaev et~al.}(2008)\citenamefont{Isaev, Prieels, and
  Keutgen}}]{Isaev-2008}
\bibinfo{author}{\bibfnamefont{S.}~\bibnamefont{Isaev}},
  \bibinfo{author}{\bibfnamefont{R.}~\bibnamefont{Prieels}}, \bibnamefont{and}
  \bibinfo{author}{\bibfnamefont{T.}~\bibnamefont{Keutgen}},
  \bibinfo{journal}{Nucl. Phys. A} \textbf{\bibinfo{volume}{809}},
  \bibinfo{pages}{1} (\bibinfo{year}{2008}).

\bibitem[{\citenamefont{Nadtochy et~al.}(2013)\citenamefont{Nadtochy, Schmitt,
  and Mazurek}}]{Nadtochy-2013}
\bibinfo{author}{\bibfnamefont{P.~N.} \bibnamefont{Nadtochy}},
  \bibinfo{author}{\bibfnamefont{C.}~\bibnamefont{Schmitt}}, \bibnamefont{and}
  \bibinfo{author}{\bibfnamefont{K.}~\bibnamefont{Mazurek}},
  \bibinfo{journal}{Phys. Scr. T} \textbf{\bibinfo{volume}{154}},
  \bibinfo{pages}{014004} (\bibinfo{year}{2013}).

\bibitem[{\citenamefont{Nix}(1969)}]{Nix-1969}
\bibinfo{author}{\bibfnamefont{J.~R.} \bibnamefont{Nix}},
  \bibinfo{journal}{Nucl. Phys. A} \textbf{\bibinfo{volume}{130}},
  \bibinfo{pages}{241} (\bibinfo{year}{1969}).

\bibitem[{\citenamefont{Prussin}(2007)}]{Prussin-book}
\bibinfo{author}{\bibfnamefont{S.~G.} \bibnamefont{Prussin}},
  \emph{\bibinfo{title}{Nuclear Physics for Applications: A Model Approach}}
  (\bibinfo{publisher}{Willey}, \bibinfo{address}{New York},
  \bibinfo{year}{2007}).

\bibitem[{\citenamefont{Kimura and Bonasera}(2005)}]{Kimura-2005z}
\bibinfo{author}{\bibfnamefont{S.}~\bibnamefont{Kimura}} \bibnamefont{and}
  \bibinfo{author}{\bibfnamefont{A.}~\bibnamefont{Bonasera}},
  \bibinfo{journal}{Phys. Rev. A} \textbf{\bibinfo{volume}{72}},
  \bibinfo{pages}{014703} (\bibinfo{year}{2005}).

\bibitem[{\citenamefont{Xia et~al.}(2014)\citenamefont{Xia, Xu, Li, and
  Shen}}]{BUU-spin-2014}
\bibinfo{author}{\bibfnamefont{Y.}~\bibnamefont{Xia}},
  \bibinfo{author}{\bibfnamefont{J.}~\bibnamefont{Xu}},
  \bibinfo{author}{\bibfnamefont{B.~A.} \bibnamefont{Li}}, \bibnamefont{and}
  \bibinfo{author}{\bibfnamefont{W.~Q.} \bibnamefont{Shen}},
  \bibinfo{journal}{Phys. Rev. C} \textbf{\bibinfo{volume}{89}},
  \bibinfo{pages}{064606} (\bibinfo{year}{2014}).

\bibitem[{\citenamefont{Chaudhuri et~al.}(2015)\citenamefont{Chaudhuri, Ghosh,
  Banerjee et~al.}}]{Chaudhuri-2015z}
\bibinfo{author}{\bibfnamefont{A.}~\bibnamefont{Chaudhuri}},
  \bibinfo{author}{\bibfnamefont{T.~K.} \bibnamefont{Ghosh}},
  \bibinfo{author}{\bibfnamefont{K.}~\bibnamefont{Banerjee}},
  \bibnamefont{et~al.}, \bibinfo{journal}{Phys. Rev. C}
  \textbf{\bibinfo{volume}{91}}, \bibinfo{pages}{044620}
  (\bibinfo{year}{2015}).

\bibitem[{\citenamefont{Mulgin}(2009)}]{Mulgin-2009}
\bibinfo{author}{\bibfnamefont{S.~I.} \bibnamefont{Mulgin}},
  \bibinfo{journal}{Nucl. Phys. A} \textbf{\bibinfo{volume}{824}},
  \bibinfo{pages}{1} (\bibinfo{year}{2009}).

\bibitem[{\citenamefont{Amorini et~al.}(2009)}]{Amorini-2009}
\bibinfo{author}{\bibfnamefont{F.}~\bibnamefont{Amorini}} \bibnamefont{et~al.},
  \bibinfo{journal}{Phys. Rev. Lett.} \textbf{\bibinfo{volume}{102}},
  \bibinfo{pages}{112701} (\bibinfo{year}{2009}).

\bibitem[{\citenamefont{Souliotis}(2010)}]{GS-CoMD-2010}
\bibinfo{author}{\bibfnamefont{G.~A.} \bibnamefont{Souliotis}},
  \bibinfo{journal}{J. Phys. CS} \textbf{\bibinfo{volume}{205}},
  \bibinfo{pages}{012019} (\bibinfo{year}{2010}).

\bibitem[{\citenamefont{Fountas et~al.}(2014)\citenamefont{Fountas, Souliotis,
  Veselsky, and Bonasera}}]{GS-RIB-2014}
\bibinfo{author}{\bibfnamefont{P.~N.} \bibnamefont{Fountas}},
  \bibinfo{author}{\bibfnamefont{G.~A.} \bibnamefont{Souliotis}},
  \bibinfo{author}{\bibfnamefont{M.}~\bibnamefont{Veselsky}}, \bibnamefont{and}
  \bibinfo{author}{\bibfnamefont{A.}~\bibnamefont{Bonasera}},
  \bibinfo{journal}{Phys. Rev. C} \textbf{\bibinfo{volume}{90}},
  \bibinfo{pages}{064613} (\bibinfo{year}{2014}).

\bibitem[{\citenamefont{Souliotis et~al.}(2014)\citenamefont{Souliotis,
  Fountas, Veselsky et~al.}}]{GS-NZ-2014}
\bibinfo{author}{\bibfnamefont{G.~A.} \bibnamefont{Souliotis}},
  \bibinfo{author}{\bibfnamefont{P.~N.} \bibnamefont{Fountas}},
  \bibinfo{author}{\bibfnamefont{M.}~\bibnamefont{Veselsky}},
  \bibnamefont{et~al.}, \bibinfo{journal}{Phys. Rev. C}
  \textbf{\bibinfo{volume}{90}}, \bibinfo{pages}{064612}
  (\bibinfo{year}{2014}).

\bibitem[{\citenamefont{Karapetyan et~al.}(2009)\citenamefont{Karapetyan,
  Balabekyan, Demekhina, and Adam}}]{Karapetyan-2009}
\bibinfo{author}{\bibfnamefont{G.~S.} \bibnamefont{Karapetyan}},
  \bibinfo{author}{\bibfnamefont{A.~R.} \bibnamefont{Balabekyan}},
  \bibinfo{author}{\bibfnamefont{N.~A.} \bibnamefont{Demekhina}},
  \bibnamefont{and} \bibinfo{author}{\bibfnamefont{J.}~\bibnamefont{Adam}},
  \bibinfo{journal}{Phys. At. Nucl.} \textbf{\bibinfo{volume}{72}},
  \bibinfo{pages}{911} (\bibinfo{year}{2009}).

\bibitem[{\citenamefont{Balabekyan et~al.}(2010)\citenamefont{Balabekyan,
  Karapetyan, Demekhina et~al.}}]{Balabekyan-2010}
\bibinfo{author}{\bibfnamefont{A.~R.} \bibnamefont{Balabekyan}},
  \bibinfo{author}{\bibfnamefont{G.~S.} \bibnamefont{Karapetyan}},
  \bibinfo{author}{\bibfnamefont{N.~A.} \bibnamefont{Demekhina}},
  \bibnamefont{et~al.}, \bibinfo{journal}{Phys. At. Nucl.}
  \textbf{\bibinfo{volume}{73}}, \bibinfo{pages}{1814} (\bibinfo{year}{2010}).

\bibitem[{\citenamefont{Deppman
  et~al.}(2013{\natexlab{a}})\citenamefont{Deppman, Andrade-II, Guimaraes
  et~al.}}]{Deppman-2013a}
\bibinfo{author}{\bibfnamefont{A.}~\bibnamefont{Deppman}},
  \bibinfo{author}{\bibfnamefont{E.}~\bibnamefont{Andrade-II}},
  \bibinfo{author}{\bibfnamefont{V.}~\bibnamefont{Guimaraes}},
  \bibnamefont{et~al.}, \bibinfo{journal}{Phys. Rev. C}
  \textbf{\bibinfo{volume}{88}}, \bibinfo{pages}{024608}
  (\bibinfo{year}{2013}{\natexlab{a}}).

\bibitem[{\citenamefont{Deppman
  et~al.}(2013{\natexlab{b}})\citenamefont{Deppman, Andrade-II, Guimaraes
  et~al.}}]{Deppman-2013b}
\bibinfo{author}{\bibfnamefont{A.}~\bibnamefont{Deppman}},
  \bibinfo{author}{\bibfnamefont{E.}~\bibnamefont{Andrade-II}},
  \bibinfo{author}{\bibfnamefont{V.}~\bibnamefont{Guimaraes}},
  \bibnamefont{et~al.}, \bibinfo{journal}{Phys. Rev. C}
  \textbf{\bibinfo{volume}{88}}, \bibinfo{pages}{064609}
  (\bibinfo{year}{2013}{\natexlab{b}}).

\bibitem[{\citenamefont{Ricciardi et~al.}(2006)\citenamefont{Ricciardi,
  Armbruster, Benlliure et~al.}}]{GSI-data-2006}
\bibinfo{author}{\bibfnamefont{M.~V.} \bibnamefont{Ricciardi}},
  \bibinfo{author}{\bibfnamefont{P.}~\bibnamefont{Armbruster}},
  \bibinfo{author}{\bibfnamefont{J.}~\bibnamefont{Benlliure}},
  \bibnamefont{et~al.}, \bibinfo{journal}{Phys. Rev. C}
  \textbf{\bibinfo{volume}{73}}, \bibinfo{pages}{014607}
  (\bibinfo{year}{2006}).

\bibitem[{\citenamefont{Schmidt et~al.}(2013)\citenamefont{Schmidt, Jurado,
  Pleskac et~al.}}]{GSI-data-2013}
\bibinfo{author}{\bibfnamefont{K.~H.} \bibnamefont{Schmidt}},
  \bibinfo{author}{\bibfnamefont{B.}~\bibnamefont{Jurado}},
  \bibinfo{author}{\bibfnamefont{R.}~\bibnamefont{Pleskac}},
  \bibnamefont{et~al.}, \bibinfo{journal}{Phys. Rev. C}
  \textbf{\bibinfo{volume}{87}}, \bibinfo{pages}{034601}
  (\bibinfo{year}{2013}).

\bibitem[{\citenamefont{Keli\'{c} et~al.}(2009)\citenamefont{Keli\'{c},
  Ricciardi, and Schmidt}}]{Spallation-2009}
\bibinfo{author}{\bibfnamefont{A.}~\bibnamefont{Keli\'{c}}},
  \bibinfo{author}{\bibfnamefont{M.~V.} \bibnamefont{Ricciardi}},
  \bibnamefont{and} \bibinfo{author}{\bibfnamefont{K.~H.}
  \bibnamefont{Schmidt}}, \bibinfo{journal}{BgNS Transactions}
  \textbf{\bibinfo{volume}{13}}, \bibinfo{pages}{98} (\bibinfo{year}{2009}).

\bibitem[{\citenamefont{Giuliani
  et~al.}(2014{\natexlab{b}})\citenamefont{Giuliani, Zheng, and
  Bonasera}}]{Bonasera-2014}
\bibinfo{author}{\bibfnamefont{G.}~\bibnamefont{Giuliani}},
  \bibinfo{author}{\bibfnamefont{H.}~\bibnamefont{Zheng}}, \bibnamefont{and}
  \bibinfo{author}{\bibfnamefont{A.}~\bibnamefont{Bonasera}},
  \bibinfo{journal}{Prog. Part. Nucl. Phys.} \textbf{\bibinfo{volume}{76}},
  \bibinfo{pages}{116} (\bibinfo{year}{2014}{\natexlab{b}}).

\bibitem[{\citenamefont{Kohley and Yennello}(2013)}]{Kohley-2014}
\bibinfo{author}{\bibfnamefont{Z.}~\bibnamefont{Kohley}} \bibnamefont{and}
  \bibinfo{author}{\bibfnamefont{S.~J.} \bibnamefont{Yennello}},
  \bibinfo{journal}{Eur. Phys. J. A} \textbf{\bibinfo{volume}{50}},
  \bibinfo{pages}{31} (\bibinfo{year}{2013}).

\bibitem[{\citenamefont{Horowitz et~al.}(2014)\citenamefont{Horowitz, Brown,
  Kim, and other}}]{Horowitz-2014}
\bibinfo{author}{\bibfnamefont{C.~J.} \bibnamefont{Horowitz}},
  \bibinfo{author}{\bibfnamefont{E.~F.} \bibnamefont{Brown}},
  \bibinfo{author}{\bibfnamefont{Y.}~\bibnamefont{Kim}}, \bibnamefont{and}
  \bibinfo{author}{\bibnamefont{other}}, \bibinfo{journal}{J. Phys. G}
  \textbf{\bibinfo{volume}{41}}, \bibinfo{pages}{093001}
  (\bibinfo{year}{2014}).

\bibitem[{\citenamefont{Yanez et~al.}(2014)\citenamefont{Yanez, Yao, King,
  Loveland, Tovesson, and Fotiades}}]{Loveland-2014}
\bibinfo{author}{\bibfnamefont{R.}~\bibnamefont{Yanez}},
  \bibinfo{author}{\bibfnamefont{L.}~\bibnamefont{Yao}},
  \bibinfo{author}{\bibfnamefont{J.}~\bibnamefont{King}},
  \bibinfo{author}{\bibfnamefont{W.}~\bibnamefont{Loveland}},
  \bibinfo{author}{\bibfnamefont{F.}~\bibnamefont{Tovesson}}, \bibnamefont{and}
  \bibinfo{author}{\bibfnamefont{N.}~\bibnamefont{Fotiades}},
  \bibinfo{journal}{Phys. Rev. C} \textbf{\bibinfo{volume}{89}},
  \bibinfo{pages}{051604} (\bibinfo{year}{2014}).

\bibitem[{\citenamefont{Hinde et~al.}(1992)\citenamefont{Hinde, Hilscher,
  Rossner et~al.}}]{Hinde-1992}
\bibinfo{author}{\bibfnamefont{D.~J.} \bibnamefont{Hinde}},
  \bibinfo{author}{\bibfnamefont{D.}~\bibnamefont{Hilscher}},
  \bibinfo{author}{\bibfnamefont{H.}~\bibnamefont{Rossner}},
  \bibnamefont{et~al.}, \bibinfo{journal}{Phys. Rev. C}
  \textbf{\bibinfo{volume}{45}}, \bibinfo{pages}{1229} (\bibinfo{year}{1992}).

\bibitem[{\citenamefont{Strecker et~al.}(1990)\citenamefont{Strecker, Wien,
  Plischke, and Scobel}}]{Strecker-1990}
\bibinfo{author}{\bibfnamefont{M.}~\bibnamefont{Strecker}},
  \bibinfo{author}{\bibfnamefont{R.}~\bibnamefont{Wien}},
  \bibinfo{author}{\bibfnamefont{P.}~\bibnamefont{Plischke}}, \bibnamefont{and}
  \bibinfo{author}{\bibfnamefont{W.}~\bibnamefont{Scobel}},
  \bibinfo{journal}{Phys. Rev. C} \textbf{\bibinfo{volume}{41}},
  \bibinfo{pages}{2172} (\bibinfo{year}{1990}).

\bibitem[{\citenamefont{Jacquet and Morjean}(2009)}]{fission-time}
\bibinfo{author}{\bibfnamefont{D.}~\bibnamefont{Jacquet}} \bibnamefont{and}
  \bibinfo{author}{\bibfnamefont{M.}~\bibnamefont{Morjean}},
  \bibinfo{journal}{Prog. Part. Nucl. Phys.} \textbf{\bibinfo{volume}{63}},
  \bibinfo{pages}{155} (\bibinfo{year}{2009}).

\bibitem[{\citenamefont{Caamano et~al.}(2013)\citenamefont{Caamano, Delaune,
  Farget et~al.}}]{Farget-2013}
\bibinfo{author}{\bibfnamefont{M.}~\bibnamefont{Caamano}},
  \bibinfo{author}{\bibfnamefont{O.}~\bibnamefont{Delaune}},
  \bibinfo{author}{\bibfnamefont{F.}~\bibnamefont{Farget}},
  \bibnamefont{et~al.}, \bibinfo{journal}{Phys. Rev. C}
  \textbf{\bibinfo{volume}{88}}, \bibinfo{pages}{024605}
  (\bibinfo{year}{2013}).

\bibitem[{\citenamefont{Rodr\'{i}guez-Tajes
  et~al.}(2014)\citenamefont{Rodr\'{i}guez-Tajes, Farget, Derkx
  et~al.}}]{Farget-2014}
\bibinfo{author}{\bibfnamefont{C.}~\bibnamefont{Rodr\'{i}guez-Tajes}},
  \bibinfo{author}{\bibfnamefont{F.}~\bibnamefont{Farget}},
  \bibinfo{author}{\bibfnamefont{X.}~\bibnamefont{Derkx}},
  \bibnamefont{et~al.}, \bibinfo{journal}{Phys. Rev. C}
  \textbf{\bibinfo{volume}{89}}, \bibinfo{pages}{024614}
  (\bibinfo{year}{2014}).

\end{thebibliography}


\newpage


\begin{figure}[htbp]                                        
\includegraphics[width=0.45\textwidth,keepaspectratio=true]{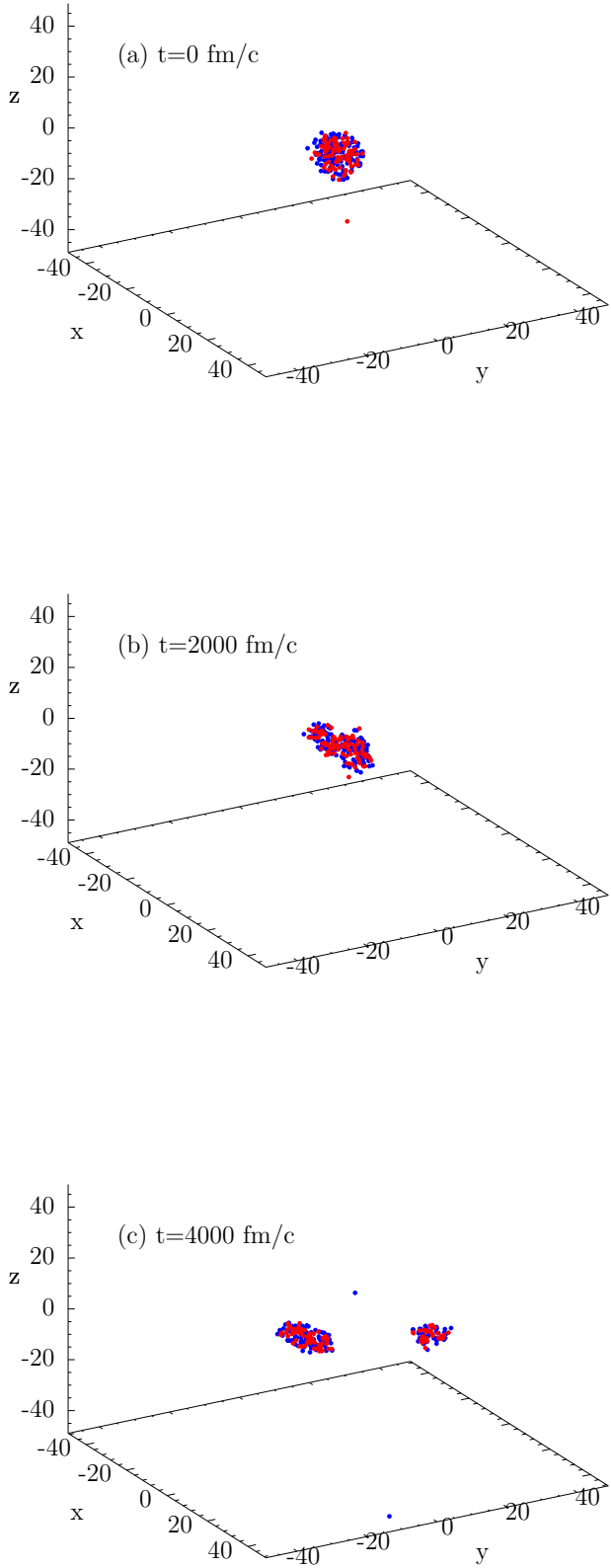}
\caption{ (Color online)
Snapshots of the CoMD predicted time evolution for the reaction p (63 MeV) + $^{232}$Th 
in the center of mass system. (Blue points: neutrons, red points: protons.)
a) t = 0 fm/c, the proton approaches the $^{232}$Th nucleus.
b) t = 2000 fm/c, the fissioning nucleus is substantially deformed (saddle configuration).
c) t = 4000 fm/c, receding fission fragments.
For this fission event, the fission time is 2500 fm/c.
}
\label{fission_panel}
\end{figure}



\begin{figure}[htbp]                                        
\includegraphics[width=0.45\textwidth,keepaspectratio=true]{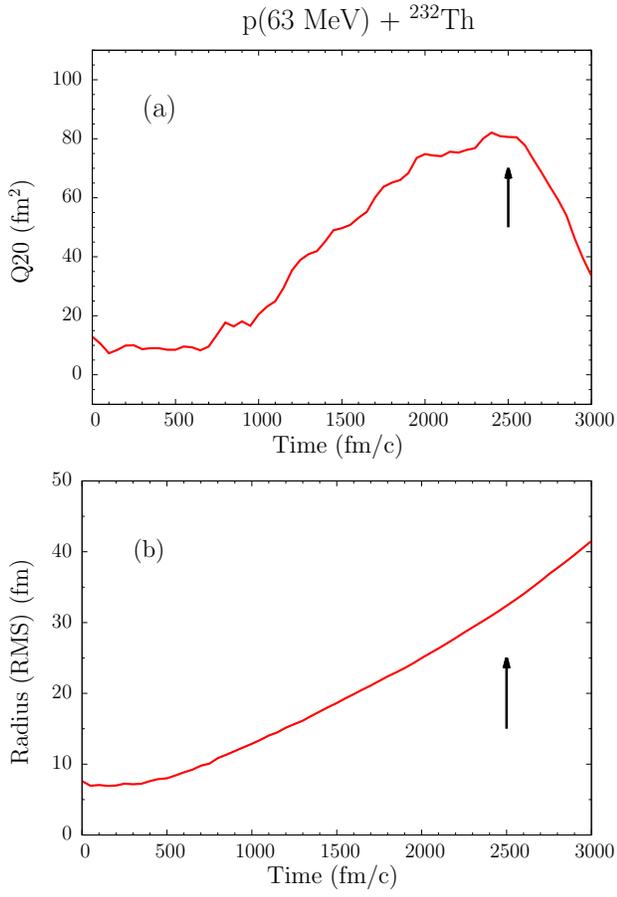}
\caption{(Color online)
CoMD calculated time evolution  of the axial quadrupole moment Q$_{20}$ (a) and  
the root-mean-square (RMS) radius (b) of the fissisoning system from the reaction
p (63 MeV) + $^{232}$Th. The arrow indicates the moment of scission. 
}
\label{fission_evolution}
\end{figure}


\begin{figure}[htbp]                                        
\includegraphics[width=0.45\textwidth,keepaspectratio=true]{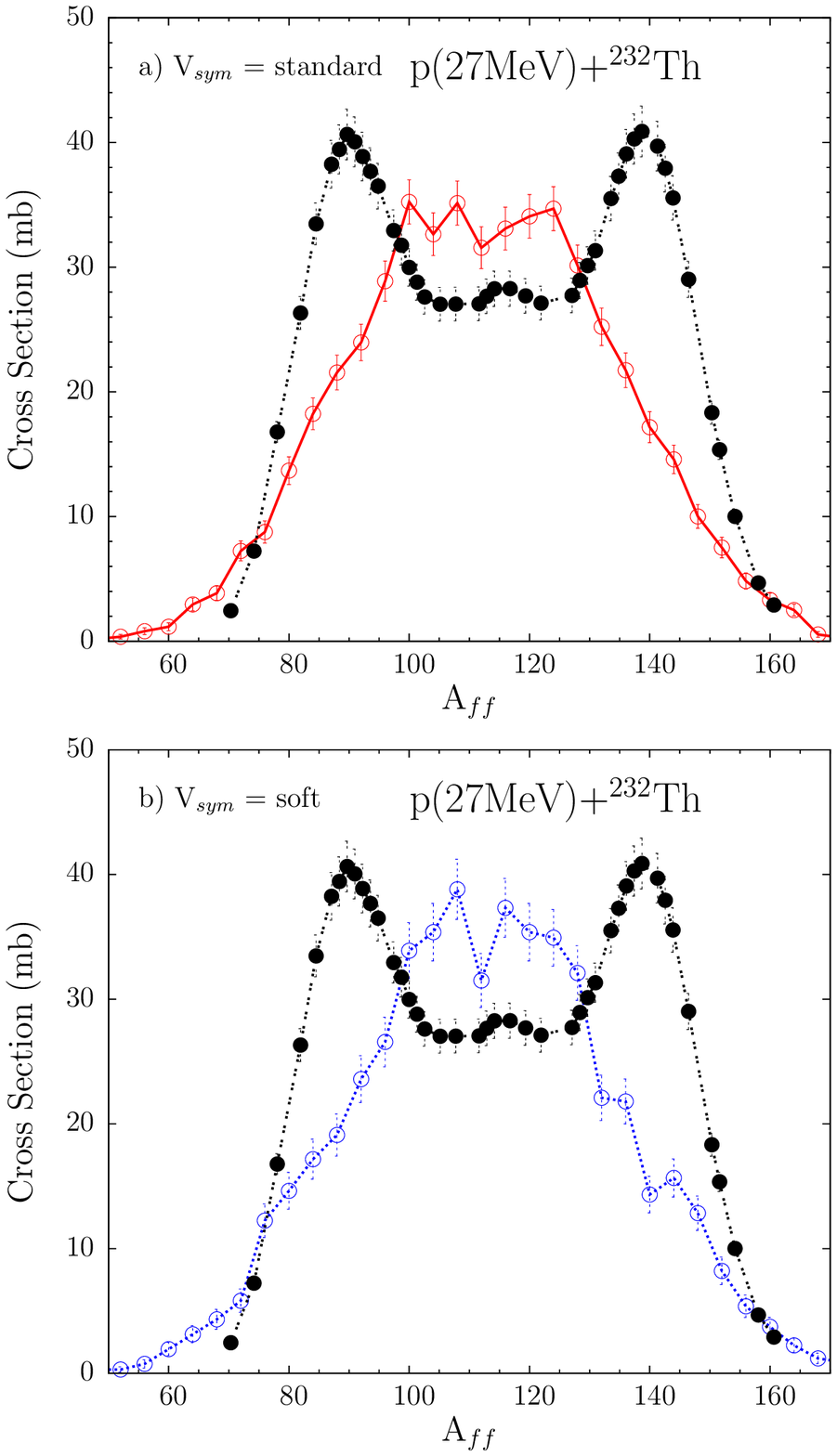}
\caption{(Color online)
a) normalized mass distributions (cross sections) of fission fragments from p (27 MeV) + $^{232}$Th. 
Full points (black): experimental data \cite{Demetriou-2010}.
Open points: CoMD calculations with the standard symmetry potential. 
b) as above, but CoMD calculations with the soft symmetry potential.
}
\label{yield_27pp232Th}
\end{figure}

\begin{figure}[htbp]                                        
\includegraphics[width=0.45\textwidth,keepaspectratio=true]{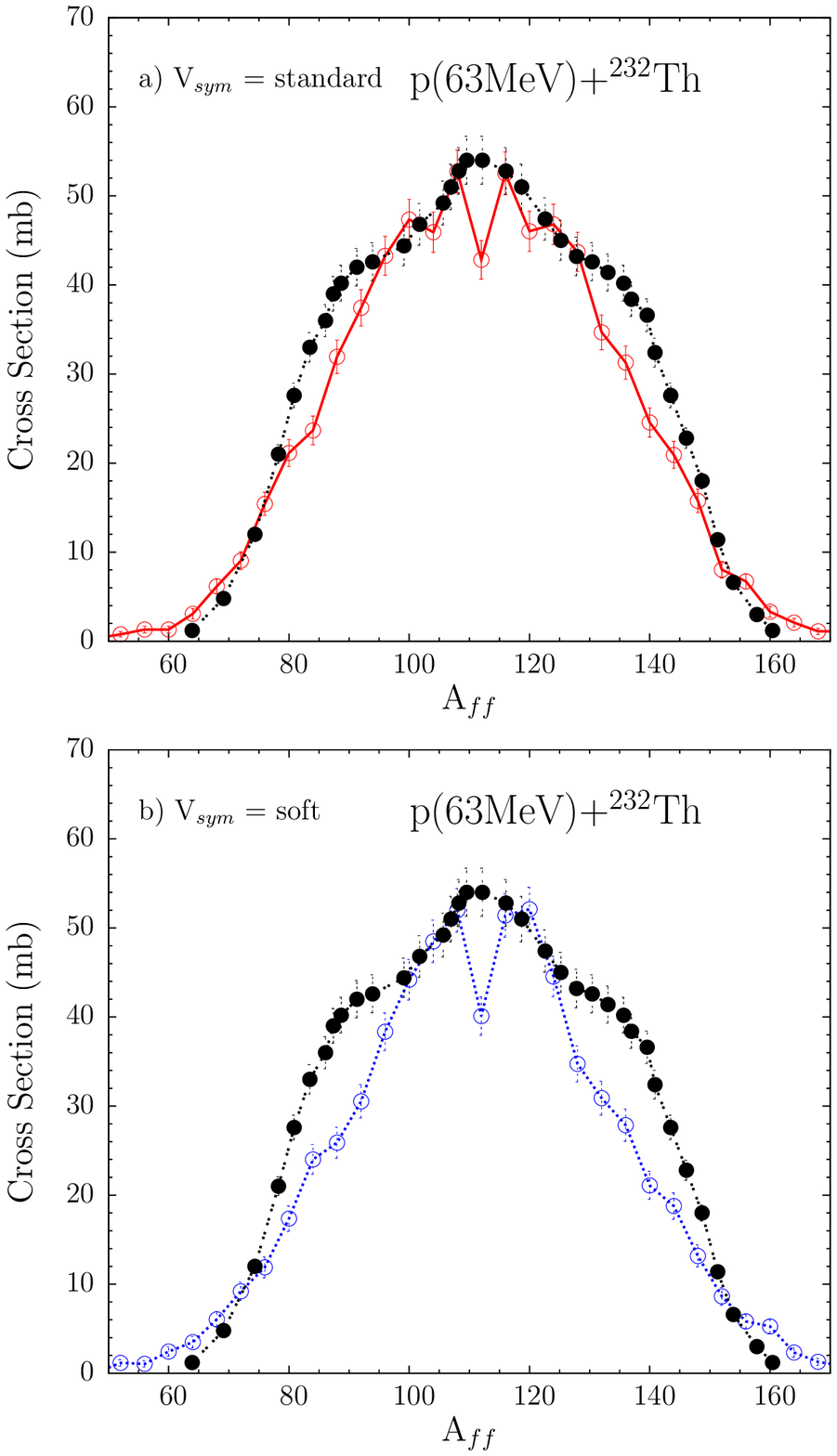}
\caption{ (Color online)
a) normalized mass distributions (cross sections) of fission fragments from p (63 MeV) + $^{232}$Th.
Full points (black): experimental data \cite{Demetriou-2010}.
Open points: CoMD calculations with the standard symmetry potential. 
b) as above, but CoMD calculations with the soft symmetry potential.
}
\label{yield_63pp232Th}
\end{figure}

\begin{figure}[htbp]                                        
\includegraphics[width=0.45\textwidth,keepaspectratio=true]{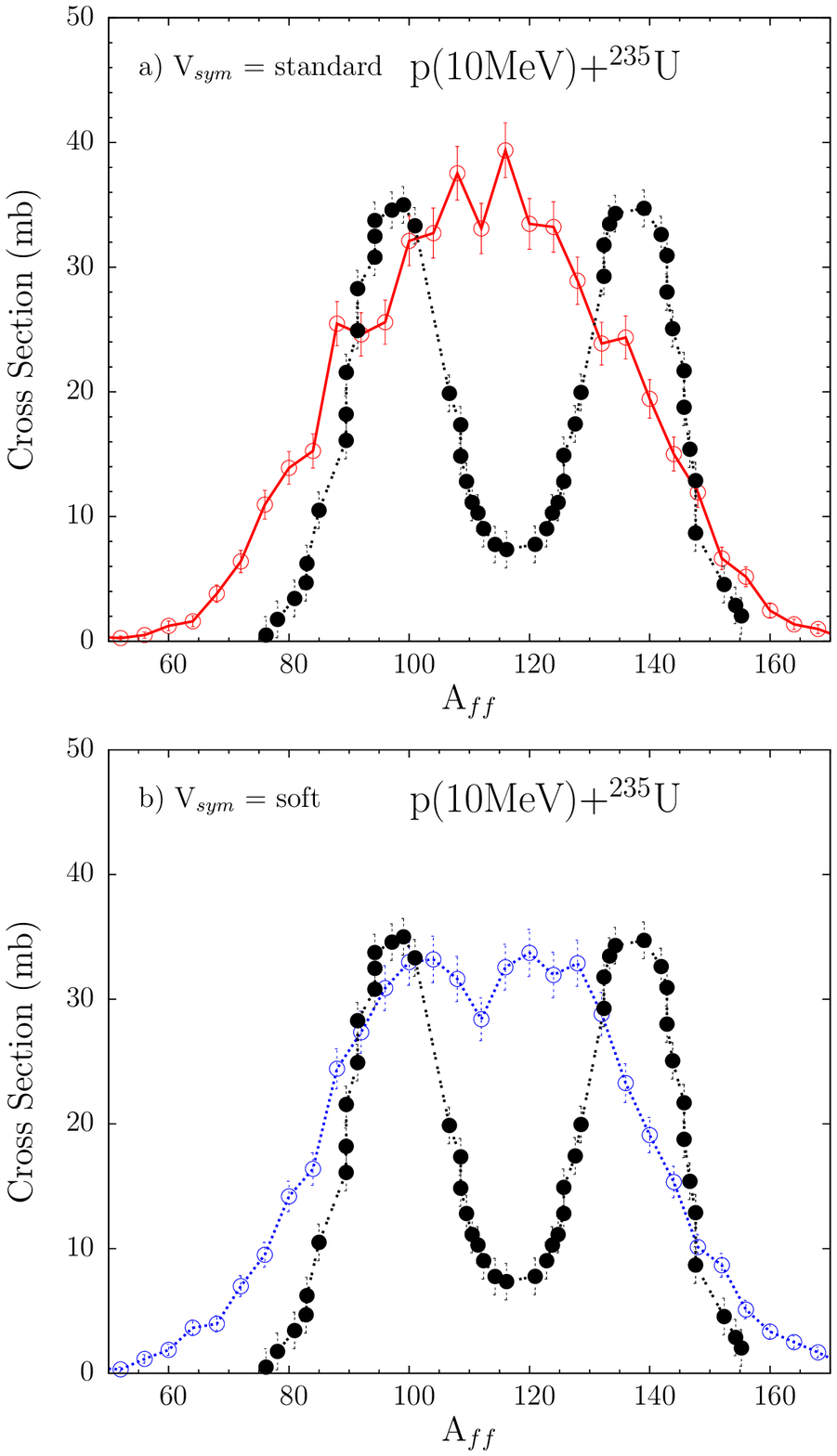}
\caption{ (Color online)
a) normalized mass distributions (cross sections) of fission fragments from p (10 MeV) + $^{235}$U. 
Full points (black): experimental data \cite{Mulgin-2009}.
Open points: CoMD calculations with the standard  symmetry potential. 
b) as above, but CoMD calculations with the soft symmetry potential.
}
\label{yield_10pp235U}
\end{figure}


\begin{figure}[htbp]                                        
\includegraphics[width=0.45\textwidth,keepaspectratio=true]{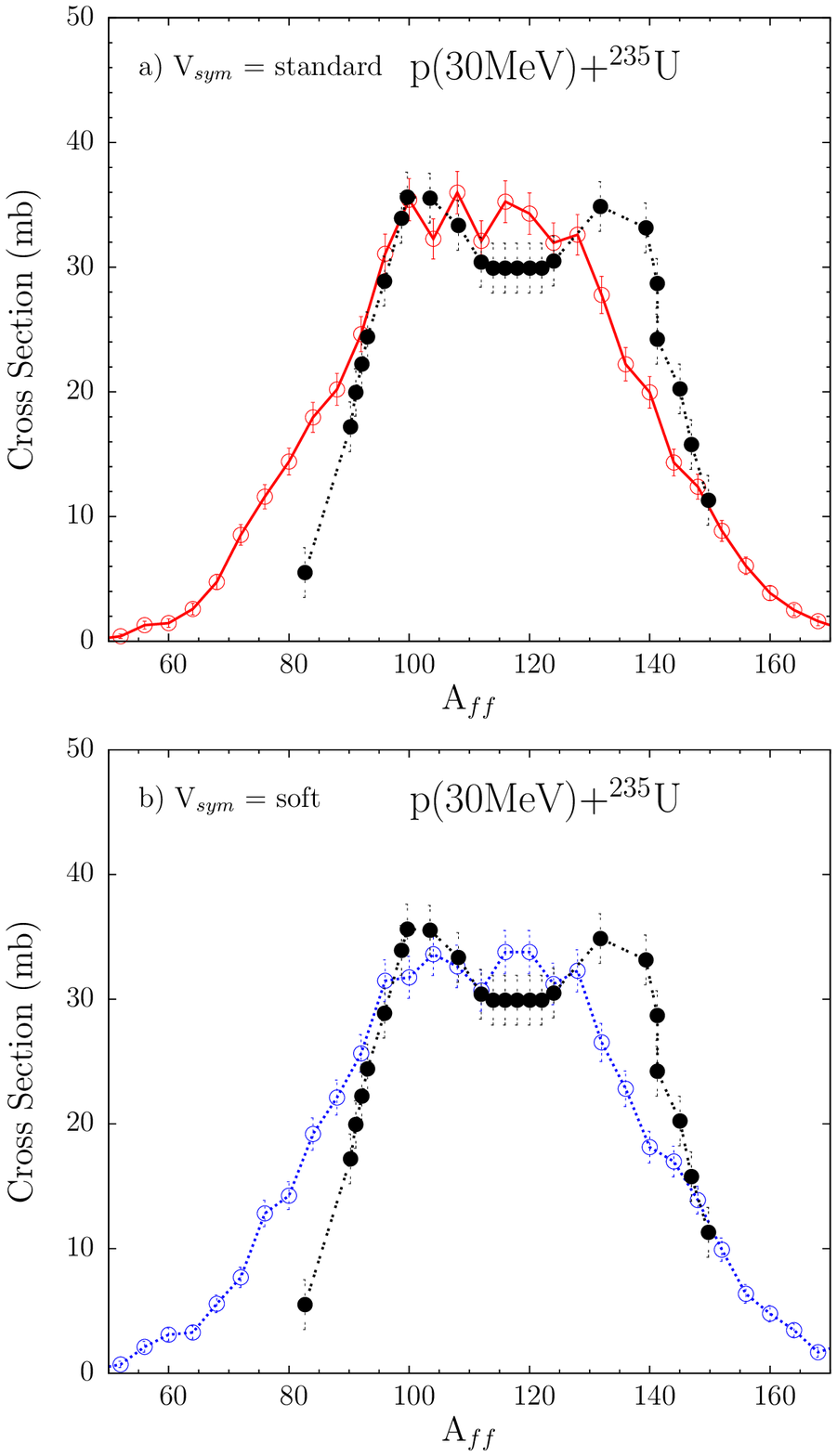}
\caption{ (Color online)
a) normalized mass distributions (cross sections) of fission fragments from p (30 MeV) + $^{235}$U. 
Full points (black): experimental data \cite{Mulgin-2009}.
Open points: CoMD calculations with the standard  symmetry potential. 
b) as above, but CoMD calculations with the soft symmetry potential.
}
\label{yield_30pp235U}
\end{figure}
\begin{figure}[htbp]                                        
\includegraphics[width=0.45\textwidth,keepaspectratio=true]{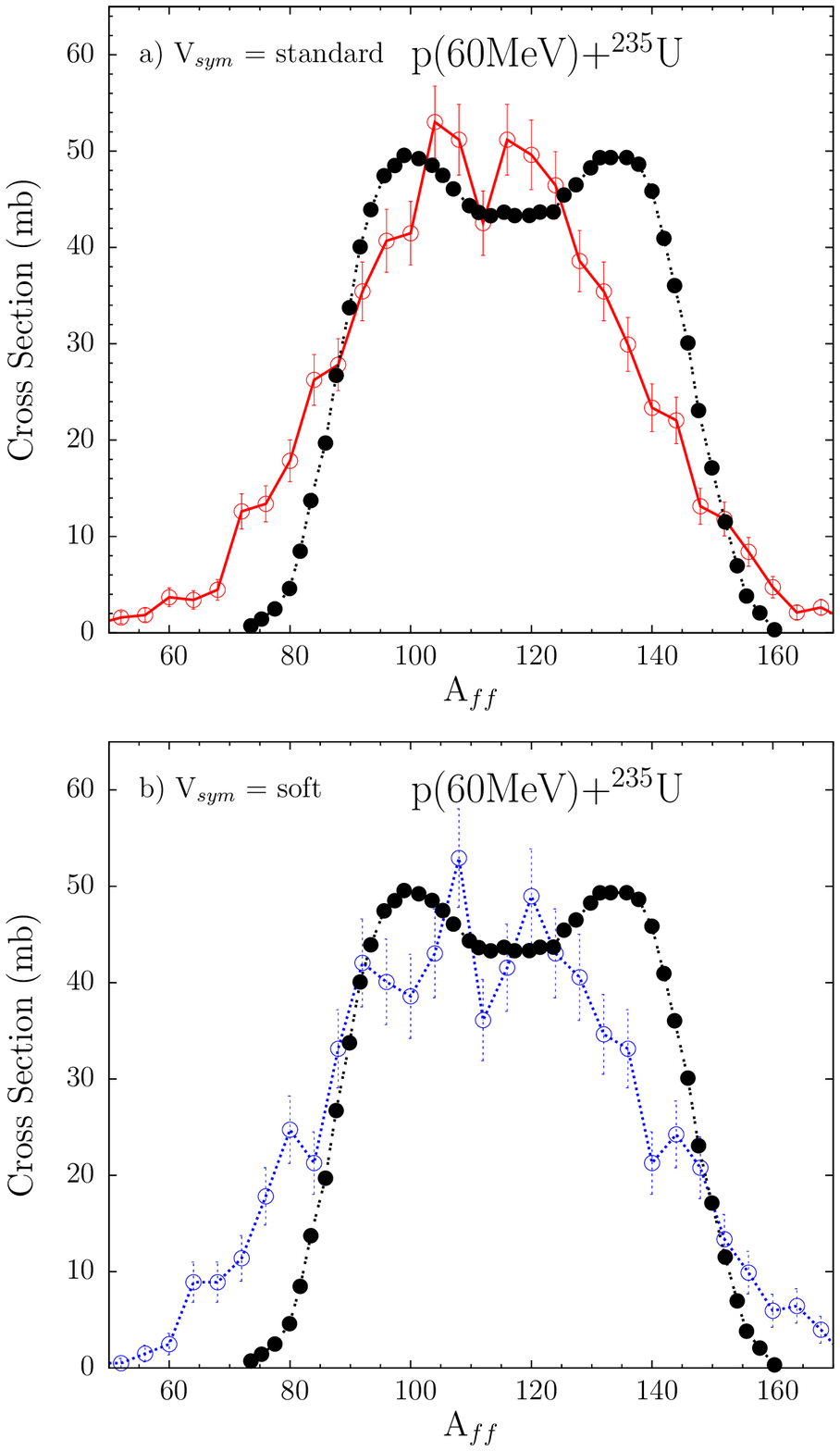}
\caption{ (Color online)
a) normalized mass distributions, cross sections of fission fragments from p (60 MeV) + U.
Full points (black): experimental data for p (60 MeV) + $^{238}$U \cite{Dui-2001}.
Open points: CoMD calculations with the standard symmetry potential for p (60 MeV) + $^{235}$U. 
b) as above, but CoMD calculations with the soft symmetry potential.
}
\label{yield_60pp235U}
\end{figure}

\begin{figure}[htbp]                                        
\includegraphics[width=0.45\textwidth,keepaspectratio=true]{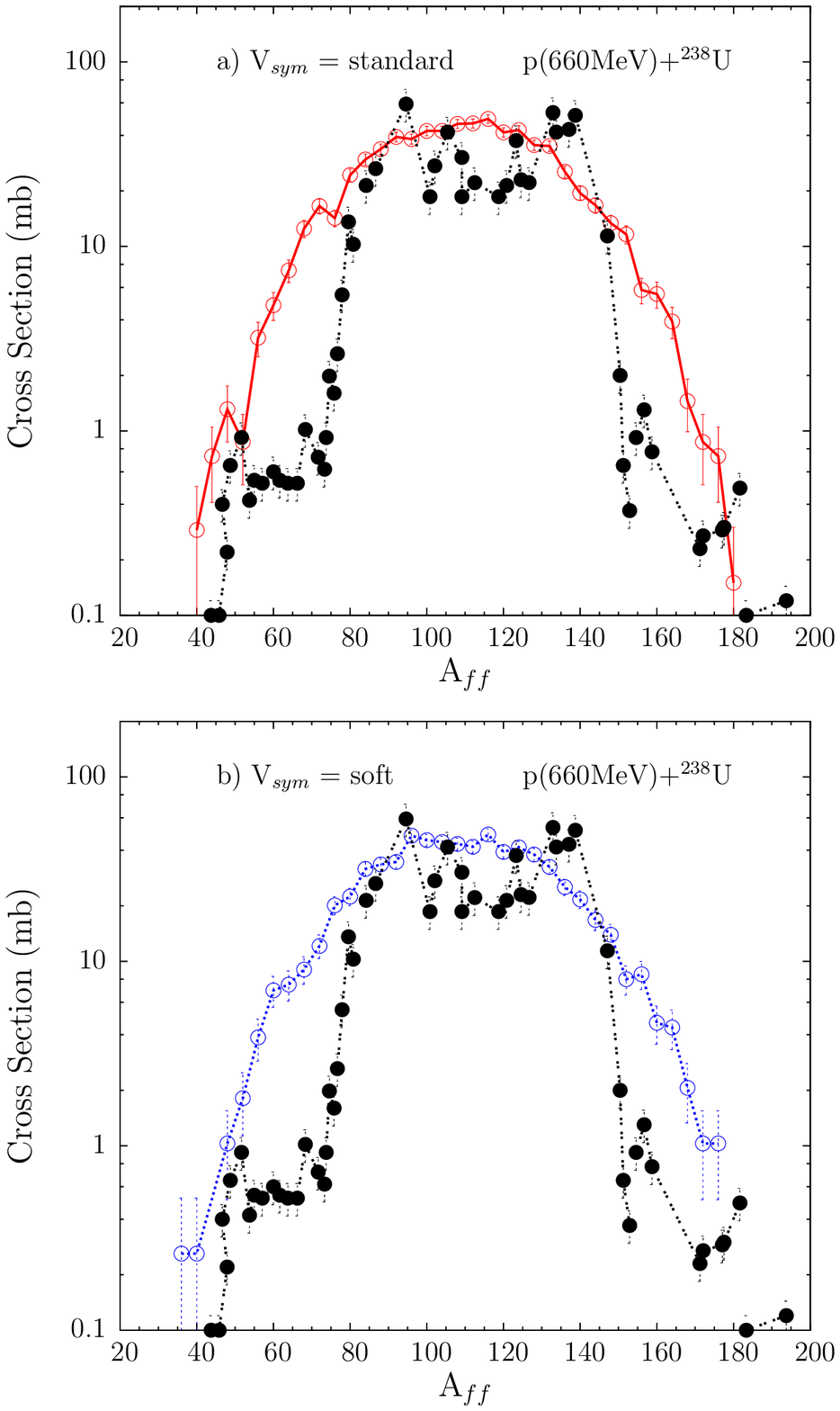}
\caption{ (Color online)
a) normalized mass distributions (cross sections) of fission fragments from p (660 MeV) + $^{238}$U.
Full points (black): experimental data \cite{Karapetyan-2009} and \cite{Balabekyan-2010}. 
Open points: CoMD calculations with the standard symmetry potential. 
b) as above, but CoMD calculations with the soft symmetry potential.
}
\label{yield_660pp238U}
\end{figure}


\begin{figure}[htbp]                                        
\includegraphics[width=0.45\textwidth,keepaspectratio=true]{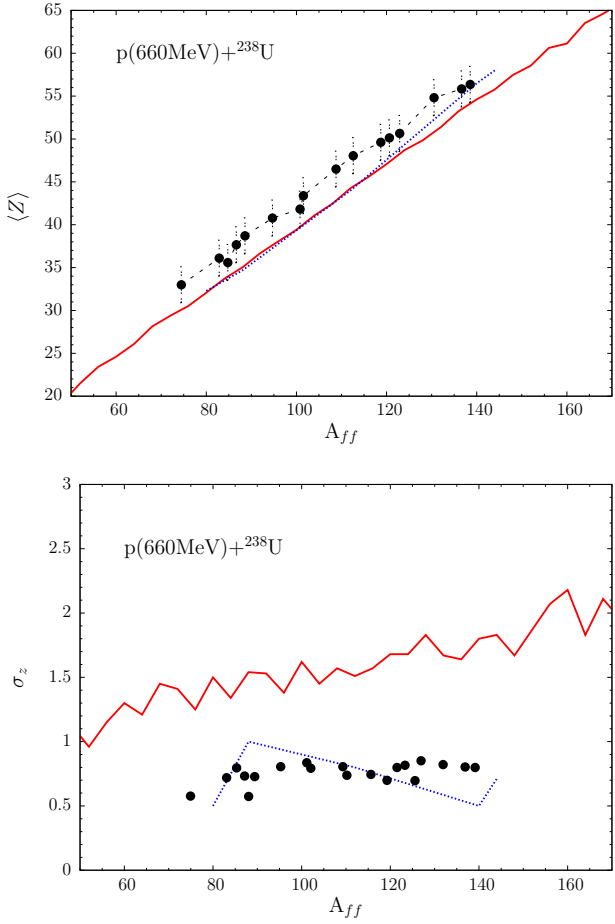}
\caption{ (Color online)
a) Fission fragment average Z vs A  from p (660 MeV) + $^{238}$U. 
Full points (black): experimental data \cite{Karapetyan-2009,Balabekyan-2010}. 
Solid (red) line: CoMD calculations 
with the standard symmetry potential. 
Dotted (blue) line : CoMD calculations 
with the additional selection of the fissioning system  not to emit any 
pre-scission protons (Z=93, see text).
b) standard deviation of the isobaric Z distribution vs A. Symbols and lines as above. 
}
\label{za_660pp238U}
\end{figure}

\begin{figure}[htbp]                                        
\includegraphics[width=0.45\textwidth,keepaspectratio=true]{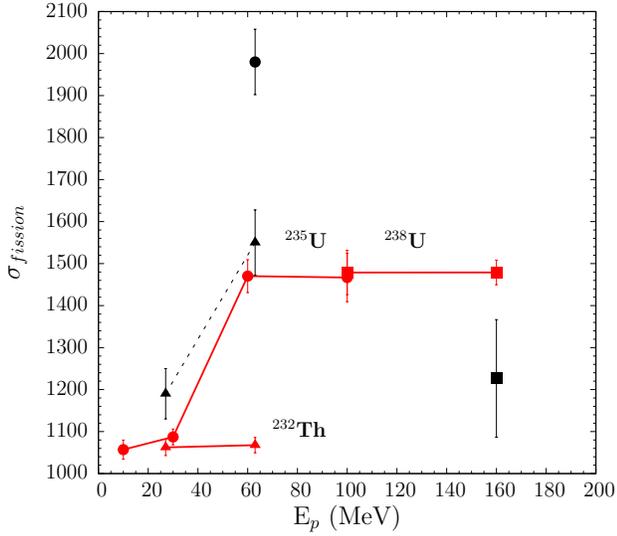}
\caption{ (Color online)
Calculated total fission cross section with respect to proton energy E$_{p}$. 
The CoMD calculations are carried out with the standard symmetry potential and are shown with 
full (red) symbols connected with full (red) lines.  The reactions are indicated as follows:
triangles: p (27, 63 MeV) + $^{232}$Th, 
circles:   p (10, 30, 60, 100  MeV) + $^{235}$U, 
squares:   p (100, 660 MeV) + $^{238}$U. 
Some experimental data are shown with closed (black) symbols as follows: 
triangles: p (27, 63 MeV) + $^{232}$Th \cite{Demetriou-2010},
circles:   p (63 MeV) + $^{238}$U  \cite{Isaev-2008},
square:   p (660 MeV) + $^{238}$U \cite{Karapetyan-2009,Balabekyan-2010}. 
The point at E$_{p}$=660 MeV is displayed at E$_{p}$=160 MeV.
The errorbars are statistical. 
}
\label{corss_sections}
\end{figure}


\begin{figure}[htbp]                                        
\includegraphics[width=0.45\textwidth,keepaspectratio=true]{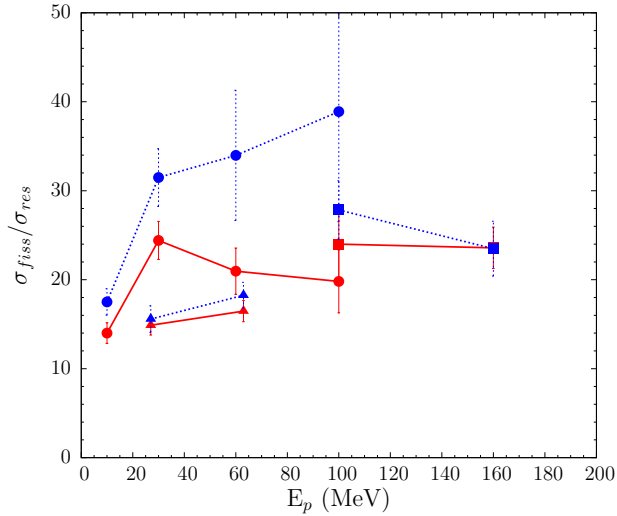}
\caption{ (Color online)
Calculated ratio of the fission cross section over residue cross section with respect to proton energy.
CoMD calculations with the standard symmetry potential are with full (red) symbols connected 
with full (red) lines. Calculations with the soft symmetry potential are with full (blue) symbols 
connected with dotted (blue) lines.  The reactions are indicated as follows:
triangles: p (27, 63 MeV) + $^{232}$Th, 
circles:   p (10, 30, 60, 100  MeV) + $^{235}$U,
squares:   p (100, 660 MeV) + $^{238}$U. 
The points at E$_{p}$=660 MeV are displayed at E$_{p}$=160 MeV.
The errorbars are statistical.
}
\label{ratio_sigma}
\end{figure}

\begin{figure}[htbp]                                        
\includegraphics[width=0.45\textwidth,keepaspectratio=true]{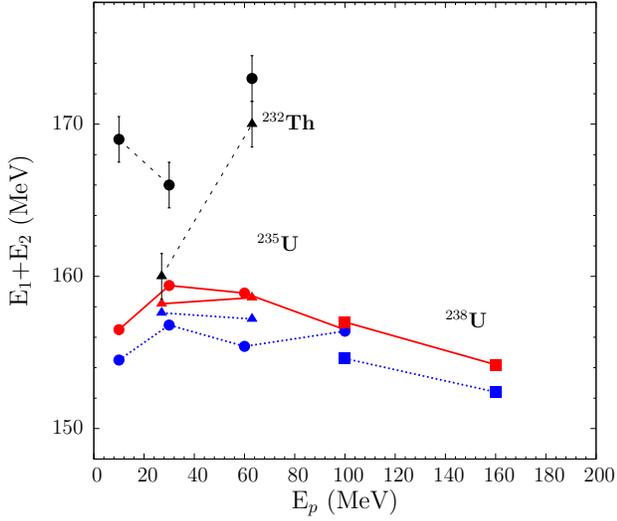}
\caption{ (Color online)
Calculated average total energy of fission fragments with respect to proton energy. 
CoMD calculations with the standard symmetry potential are with full (red) symbols connected 
with full (red) lines. Calculations with the soft symmetry potential are with full (blue) symbols 
connected with dotted (blue) lines.  The reactions are indicated as follows:
triangles: p (27, 63 MeV) + $^{232}$Th, 
circles:   p (10, 30, 60, 100  MeV) + $^{235}$U, 
squares:   p (100, 660 MeV) + $^{238}$U. 
Some experimental data are shown with closed (black) symbols as follows: 
triangles: p (27, 63 MeV) + $^{232}$Th \cite{Demetriou-2010}, 
circles:   p (10, 30 MeV) + $^{235}$U  \cite{Mulgin-2009} and
p (63 MeV) + $^{238}$U  \cite{Isaev-2008}.
The points at E$_{p}$=660 MeV are displayed at E$_{p}$=160 MeV.
(Statistical errorbars on the theoretical points are about 
1.0-1.5 MeV and are not shown for clarity.)
}
\label{total_energy}
\end{figure}
\begin{figure}[htbp]                                        
\includegraphics[width=0.45\textwidth,keepaspectratio=true]{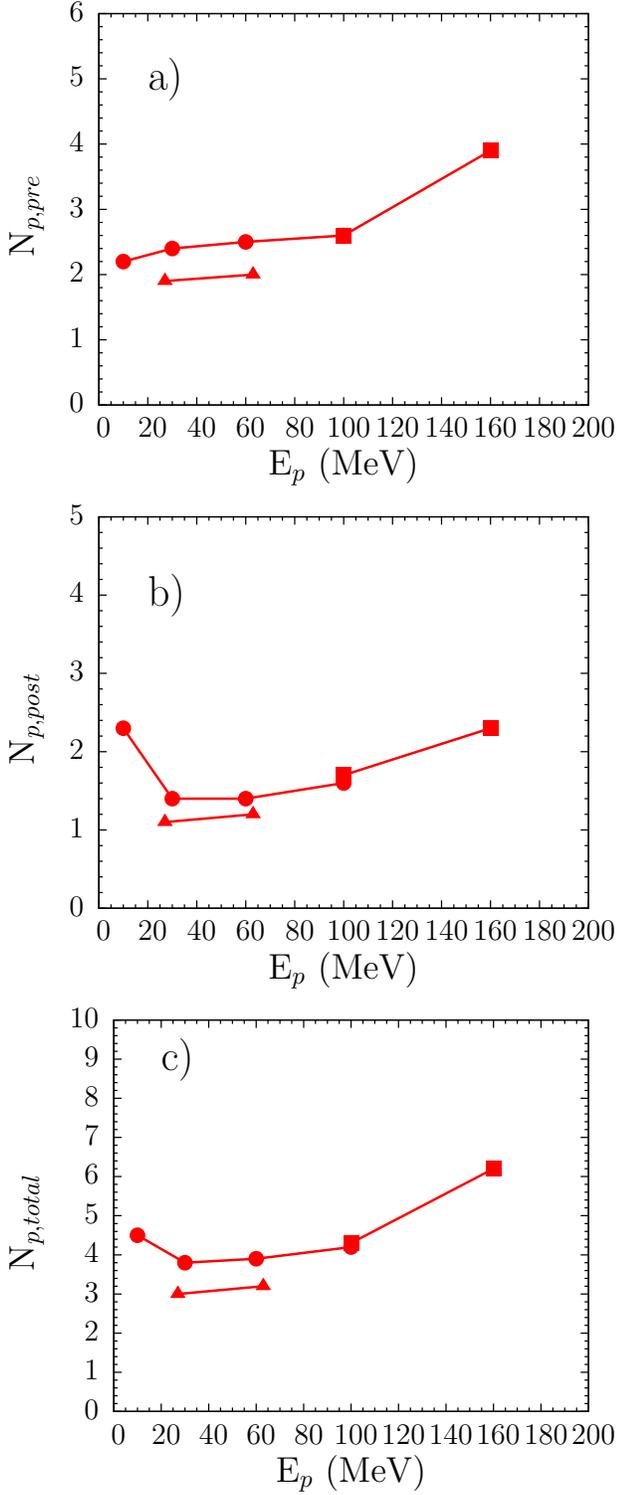}
\caption{ (Color online)
Calculated average proton multiplicities with respect to incident proton energy: 
a) pre-scission b) post-scission and c) total multiplicities.
CoMD calculations are carried with the standard symmetry potential.
The reactions are indicated as follows:
triangles: p (27, 63 MeV) + $^{232}$Th, 
circles:   p (10, 30, 60, 100  MeV) + $^{235}$U, 
squares:   p (100, 660 MeV) + $^{238}$U. 
The point at E$_{p}$=660 MeV is displayed at E$_{p}$=160 MeV.
}
\label{proton_multiplicities}
\end{figure}
\begin{figure}[htbp]                                        
\includegraphics[width=0.45\textwidth,keepaspectratio=true]{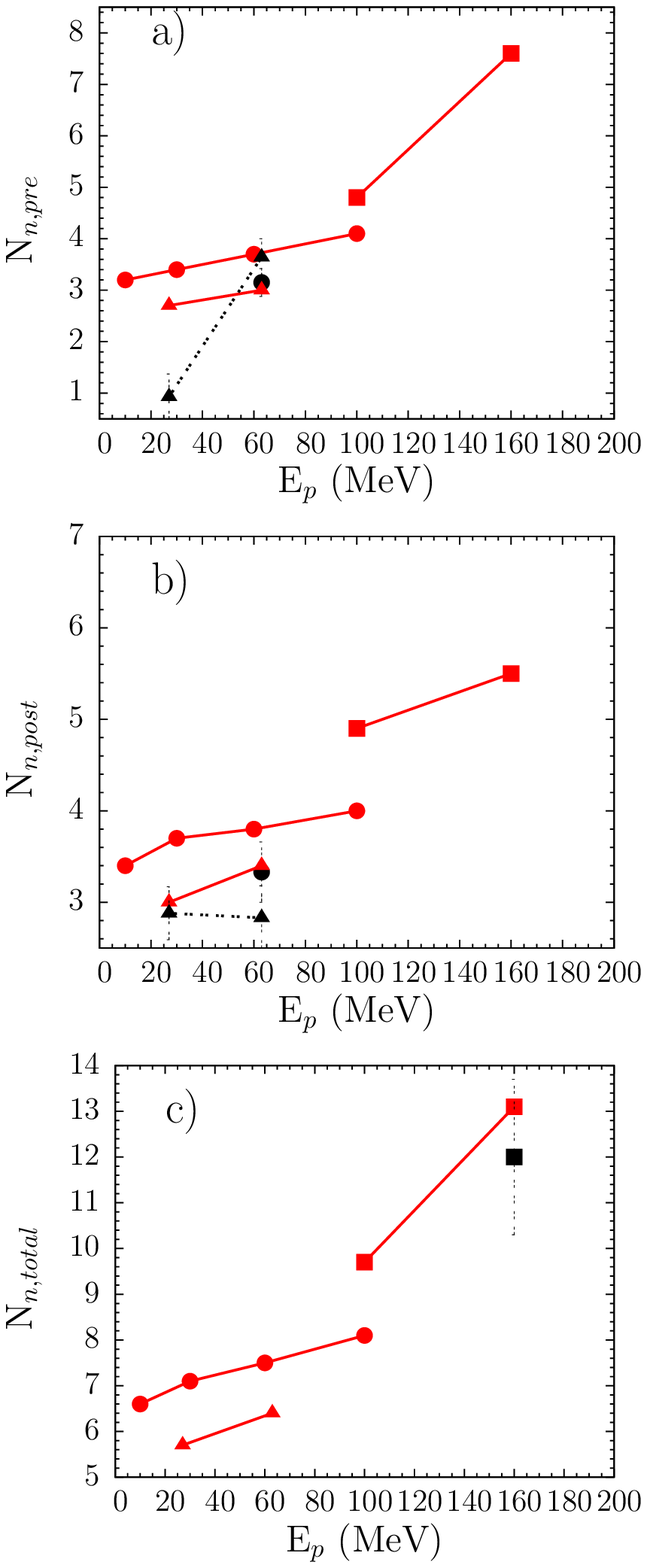}
\caption{ (Color online)
Calculated average neutron multiplicities with respect to incident proton energy: 
a) pre-scission b) post-scission and c) total multiplicities.
CoMD calculations are carried with the standard symmetry potential.
The reactions are indicated as follows:
triangles: p (27, 63 MeV) + $^{232}$Th, 
circles:   p (10, 30, 60, 100  MeV) + $^{235}$U, 
squares:   p (100, 660 MeV) + $^{238}$U. 
Some experimentail data are shown with closed (black) symbols  as follows: 
triangles: p (27, 63 MeV) + $^{232}$Th \cite{Demetriou-2010},
circle:    p (63 MeV) + $^{238}$U  \cite{Demetriou-2010}.    
square:    p (660 MeV) + $^{238}$U  \cite{Deppman-2013a}.
The point at E$_{p}$=660 MeV is displayed at E$_{p}$=160 MeV.
}
\label{neutron_multiplicity}
\end{figure}

\begin{figure}[htbp]                                        
\includegraphics[width=0.45\textwidth,keepaspectratio=true]{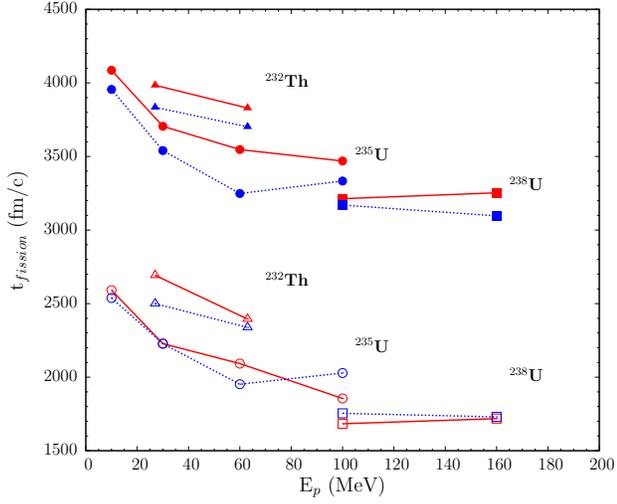}
\caption{ (Color online)
Calculated fission time with respect to incident proton energy.
CoMD calculations with the standard symmetry potential are with (red) symbols connected 
with full (red) lilnes. Calculations with the soft symmetry potential are with (blue) symbols 
connected with dotted (blue) lines.  
The full symbols (upper half of the figure) are with the full ensemble of the fissioning nuclei,
whereas the open symbols (lower half) are with the selection of the fissioning system  not to emit any 
pre-scission protons (see text).
The reactions are indicated as follows:
triangles: p (27, 63 MeV) + $^{232}$Th, 
circles:   p (10, 30, 60, 100  MeV) + $^{235}$U,
squares:   p (100, 660 MeV) + $^{238}$U. 
The points at E$_{p}$=660 MeV are displayed at E$_{p}$=160 MeV.
}
\label{fission_time}
\end{figure}

\begin{figure}[htbp]                                        
\includegraphics[width=0.45\textwidth,keepaspectratio=true]{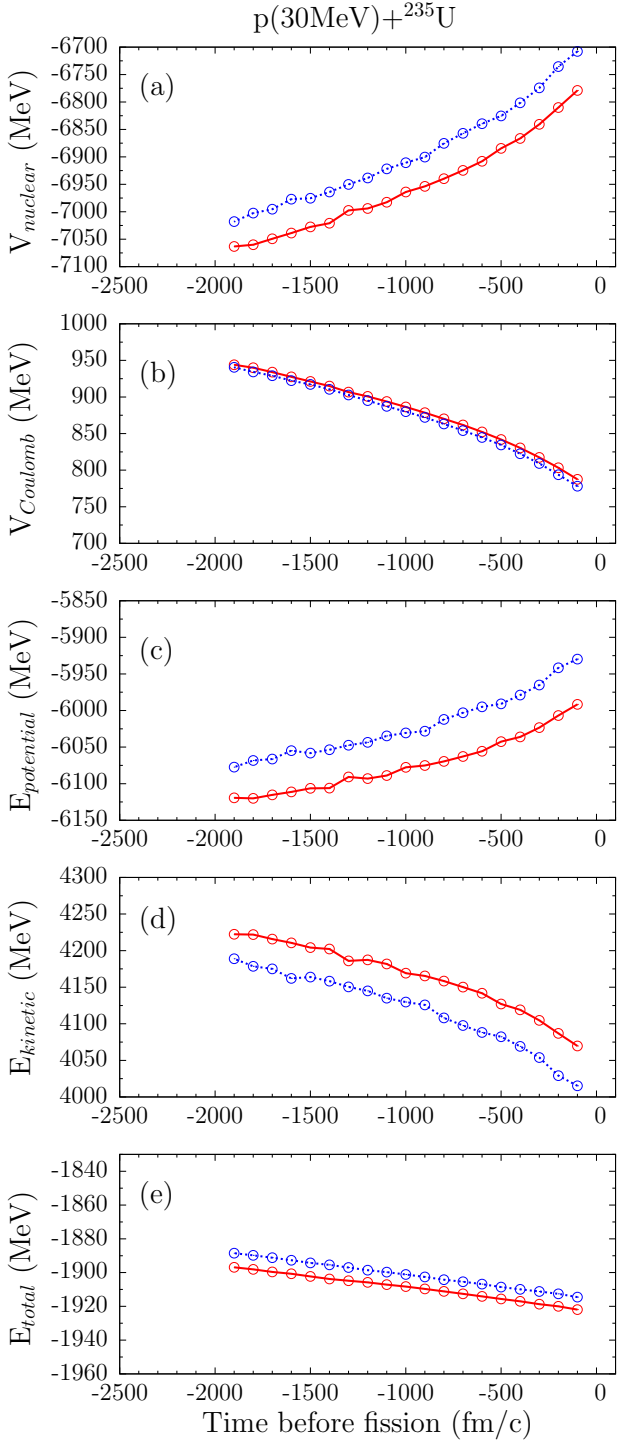}
\caption{(Color online)
Evolution of ensemble average CoMD energies of fissioning nuclei in the interaction 
of p(30 MeV) with $^{235}$U. The moment of scission is taken as t=0 fm/c.
(Red) points connected with solid lines are with the standard symmetry potential. 
(Blue) points connected with dotted lines are with 
the soft symmetry potential. The energies are: a) Nuclear potential energy (the sum of the two-body 
interaction energy, the three-body interaction energy, the surface energy and the symmetry energy).
b) Coulomb potential energy. c) Total potential energy [sum of energies of (a) and (b)]. 
d) Kinetic energy. e) Total energy [sum of (c) and (d)].
}
\label{energy_evolution_1}
\end{figure}

\begin{figure}[htbp]                                        
\includegraphics[width=0.45\textwidth,keepaspectratio=true]{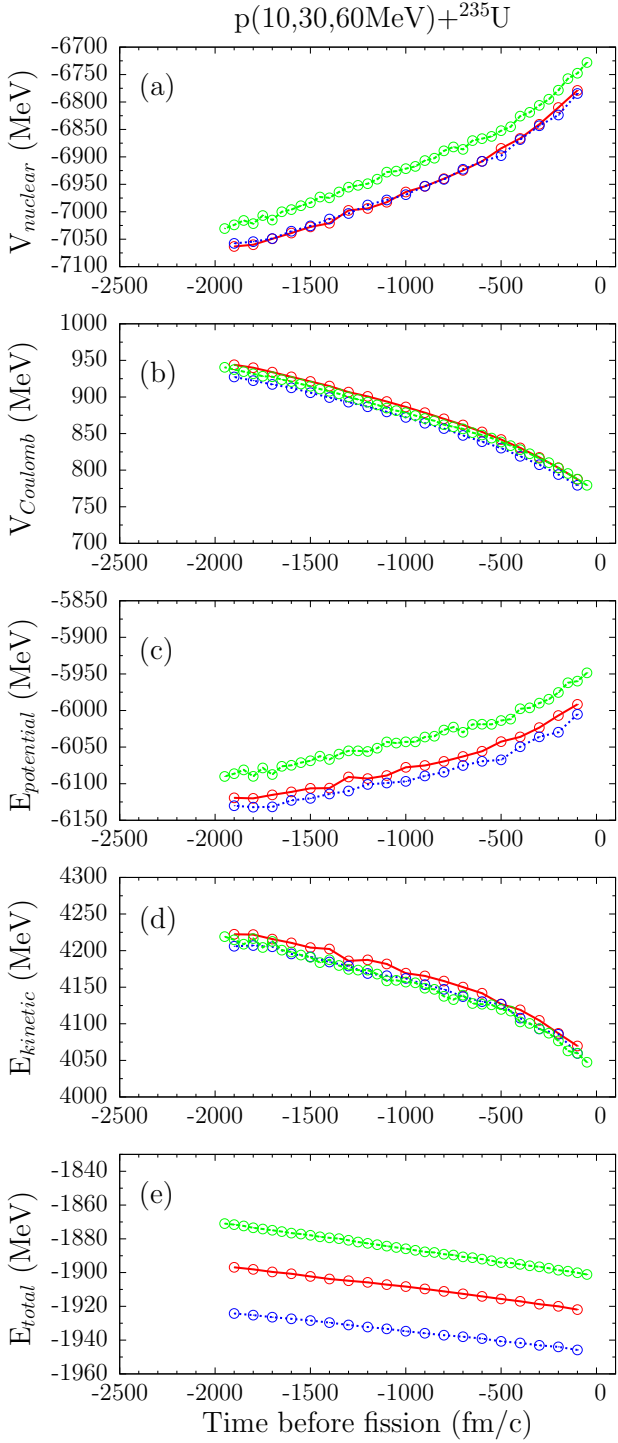}
\caption{(Color online)
Evolution of ensemble average CoMD energies of fissioning nuclei in the interaction 
of p with $^{235}$U. 
The reactions are with: 10 MeV protons: (blue)   points connected with dotted line,
                        30 MeV protons: (red)    points connected with solid line,
                        60 MeV protons: (green)  points connected with dashed line.
The moment of scission is taken as t=0 fm/c.
The calculations are with the standard symmetry potential. 
The energies are: a) Nuclear potential energy (the sum of the two-body 
interaction energy, the three-body interaction energy, the surface energy and the symmetry energy).
b) Coulomb potential energy. c) Total potential energy [sum of energies of (a) and (b)].
d) Kinetic energy. e) Total energy [sum of (c) and (d)].
}
\label{energy_evolution_2}
\end{figure}


\end{document}